%% file: draft_nstar_v2.tex
\documentclass[
reprint,
amsmath,amssymb,
]{revtex4-2}
\usepackage{soul}
\usepackage[utf8]{inputenc}

\usepackage{graphicx}
\usepackage{dcolumn}
\usepackage{bm}
\usepackage{array} 
\usepackage{mhchem} 

\usepackage{hyperref}
\hypersetup{
    colorlinks,
    citecolor=blue,
    filecolor=blue,
    linkcolor=blue,
    urlcolor=blue
}

\usepackage[mathlines]{lineno}
\usepackage{xcolor}

\begin{document}

\title{Direct Neutron Reactions in Storage Rings Utilizing a Supercompact Cyclotron Neutron Target}


\author{Ariel Tarifeño-Saldivia}
 \email{atarisal@ific.uv.es}
 \affiliation{Instituto de Física Corpuscular (CSIC-Universitat de Val\`encia), Valencia, Spain}
\author{Iris Dillmann}%
\affiliation{TRIUMF, Vancouver BC, Canada}
\affiliation{Department of Physics and Astronomy, University of Victoria, Victoria BC, Canada
}%
\author{César Domingo-Pardo}
\affiliation{Instituto de Física Corpuscular (CSIC-Universitat de Val\`encia), Valencia, Spain}
\author{Yuri A. Litvinov}
\affiliation{GSI Helmholtzzentrum f\"ur Schwerionenforschung GmbH, Darmstadt, Germany}
\affiliation{Institut f\"ur Kernphysik, Universit\"at zu K\"oln,  K\"oln, Germany}%

\date{\today}

\begin{abstract}
We propose a new approach for a high-density free-neutron target, primarily aimed at nuclear astrophysics reaction studies in inverse kinematics with radioactive ions circulating in a storage ring. The target concept integrates four key subsystems: a neutron production source driven by a supercompact cyclotron utilizing $^9$Be($p,xn$) reactions, an optimized moderator/reflector assembly using either heavy water or beryllium oxide with a graphite reflector shell to thermalize fast neutrons, a cryogenic liquid hydrogen moderator to maximize thermal neutron density in the interaction region, and beam pipe geometries that enable neutron-ion interactions while maintaining vacuum conditions for ion circulation. This integrated approach focuses on the feasibility by incorporating readily available technologies. Using a commercial supercompact cyclotron delivering a proton beam of 130 $\mu$A, the design achieves thermal neutron areal densities of $\sim3.4\times10^{6}$\,n/cm$^2$ for a proof-of-concept demonstrator at the CRYRING ion-storage ring at GSI Darmstadt. This autonomous accelerator-target assembly design enables deployment at both, in-flight and ISOL facilities, to exploit their complementary production mechanisms.

Potential upgrades based on higher-energy and/or higher-current cyclotrons will enable an increase in areal density to $\sim$10$^9$ n/cm$^2$. In combination with a customized low-energy storage ring and a radioactive ion-beam facility, the proposed solution could deliver luminosities above 10$^{23}$ cm$^{-2}$ s$^{-1}$, thereby enabling neutron capture measurements of $\sim$mb cross sections within a few days of experiment. 
The proposed system represents a significant milestone towards enabling large neutron-capture surveys on short-lived nuclei, thereby opening a new avenue for understanding the synthesis of heavy elements in our universe. \\
\\
\textit{Accepted for publication in Physical Review Accelerators and Beams. Selected as an Editors' Suggestion.}

\end{abstract}



\maketitle

\tableofcontents

\section{\label{sec:intro}Introduction}
Neutron-capture reactions drive the nucleosynthesis of heavy elements in stars and represent one of the main nuclear physics inputs for astrophysical models and for understanding the origin and formation of the chemical elements in the universe\,\cite{BBFH}.
After more than 70 years of experimental developments, practically 95\% of the neutron capture cross sections of stable isotopes have been measured in the laboratory, but only a few cross sections of radioactive isotopes have been experimentally determined\,\cite{Kaeppeler11,Dillmann23}. This is due to the challenge of producing samples with a sufficiently large number of radioactive atoms and limitations in present techniques for measuring neutron-capture cross sections\,\cite{Domingo23b}. Most of the remaining shorter-lived slow neutron capture ($s$) process branching nuclei are out of reach with state-of-the-art activation and time-of-flight techniques\,\cite{Domingo25}. Additionally, stellar explosive environments, like the rapid ($r$) neutron capture process, involve more than 5000 radioactive isotopes and thus nucleosynthesis calculations need to rely mostly on statistical theoretical models\,\cite{Arnould07,Panov10,Thielemann11,Arcones23,Rauscher22,Goriely23}. However, uncertainties of theoretical neutron-capture cross sections outside the valley of stability can reach orders of magnitude\,\cite{Liddick2016, Martinet24}. 

A breakthrough in this field could be therefore accomplished if one could directly measure the neutron-induced reactions in the involved radioactive species, something which is not feasible with fixed-target experiments (direct kinematics), like neutron time-of-flight or neutron-activation experiments.
A new approach for measuring neutron-induced reactions on radioactive ions in inverse kinematics was proposed in Ref.\,\cite{Reifarth14}. 
This proposal included a radioactive ion-beam facility (RIB), an ion storage ring, and a neutron target based on a high neutron flux fission reactor. However, using a nuclear reactor for this application was found rather difficult and an alternative approach was proposed later using a neutron-spallation source\,\cite{Reifarth17}. Several realization options were discussed in the latter work, in particular utilizing readily available spallation neutron sources like the one at Los Alamos National Laboratory\,\cite{Nowicki17} or the one at CERN n\_TOF\,\cite{Esposito23}. 

Nevertheless, none of these two spallation facilities also hosts the other two sub-facilities required: the RIB facility and the ion storage ring. In addition to the storage ring, the neutron target, and the RIB facility, a suitable method for detecting the neutron-capture products is required. A proof-of-concept project to demonstrate the attainable neutron density via the spallation approach and perform one-pass experiments with stable ions is presently being pursued at Los Alamos\,\cite{Mosby17,Cooper24}. 

In this article we explore the feasibility of replacing the complex and resource-intensive nuclear reactors or spallation neutron sources with a significantly simpler alternative: a heavily moderated neutron source driven by a supercompact cyclotron. While numerous compact accelerator-driven neutron sources (CANS) already exist\,\cite{Lavelle08,Anderson16,Carpenter19,Brueckel20,CANS_IAEA}, a comparison of current and proposed CANS designs with present and future storage-ring configurations\,\cite{Grieser12,Catherall19,Dillmann23} reveals that no existing CANS meets (yet) the requirements for experiments envisioned at a storage ring.

Therefore, the design of a suitable neutron target should, from the beginning, account for its integration requirements and constraints within a storage-ring facility. The primary motivation behind this new target proposal is therefore to ensure feasibility, while allowing for progressive optimization of performance and scalability in subsequent upgrades, ultimately aiming at performance levels comparable to previous approaches \cite{Reifarth17}.

The concept of the new neutron target is introduced in Section~\ref{sec:conc_des}, along with results from thorough Monte Carlo (MC) simulations and a description of the optimization strategy for the compact target design. Section~\ref{sec:implementation} outlines the  implementation of a demonstrator of the neutron target and possible upgrades, primarily involving higher-current cyclotrons. In Section \ref{sec:CRYRING}, we present a proof-of-principle experiment utilizing the existing GSI facility. Section~\ref{sec:future} discusses potential optimizations and their implementation in future facilities, while Section~\ref{sec:astro} describes representative astrophysical scenarios that could be explored using the proposed developments. Finally, Section~\ref{sec:summary} summarizes the main conclusions and offers a forward-looking perspective, inspired by the advancement of super-compact superconducting cyclotrons.

\section{\label{sec:conc_des} Conceptual design of the neutron target}
The design of a neutron target for storage rings must balance competing constraints from multiple subsystems. The storage ring geometry limits the available space (typically 2-4~m straight sections) and requires compatibility with ultra-high vacuum systems. The characteristics of the ion beam (revolution frequencies of 100-200 kHz and circulating intensities of 10$^7$-10$^9$ particles) define the luminosity requirements. Commercially available cyclotron technology  provides beam currents from 130 $\mu$A to 1.6 mA. These constraints define a target areal density requirement of 10$^6$-10$^9$\,n/cm$^2$ depending on facility type and scientific goals. We develop an optimization methodology to achieve maximum thermal neutron density within these constraints, following a three-stage approach of increasing complexity: single-material moderator studies (Sec.\,\ref{sec:SingleMod}), moderator/reflector combinations (Sec.\,\ref{sec:ModRef}), and cryogenic enhancement (Sec.\,\ref{sec:CryoMod}). The resulting target performance is then evaluated for three reference scenarios: a proof-of-concept at CRYRING@GSI (Sec.\,\ref{sec:CRYRING}), and future dedicated facilities at CERN-ISOLDE and TRIUMF-ISAC (Sec.\,\ref{sec:future}).

For particle beam energies in the keV range or higher, it is highly desirable that the target neutrons have thermal or cold energies, E$<$25~meV. This energy selection is crucial for minimizing systematic uncertainties during the analysis of inverse reaction kinematics, as it simplifies the energy and momentum considerations of the interaction. Consequently, our conceptual design for this target maximizes  the fraction of neutrons within the thermal or cold energy range in the designated target area. The proposed design comprises four integrated subsystems: (1) a storage-ring beam pipe defining the neutron interaction volume, (2) a cyclotron beam pipe housing a beryllium target for the neutron source, (3) a moderator/reflector assembly embedding both beam pipes, and (4) a cryogenic moderator enhancement around the ion beam path. 

\begin{figure}
    \resizebox{0.5\textwidth}{!}{\includegraphics{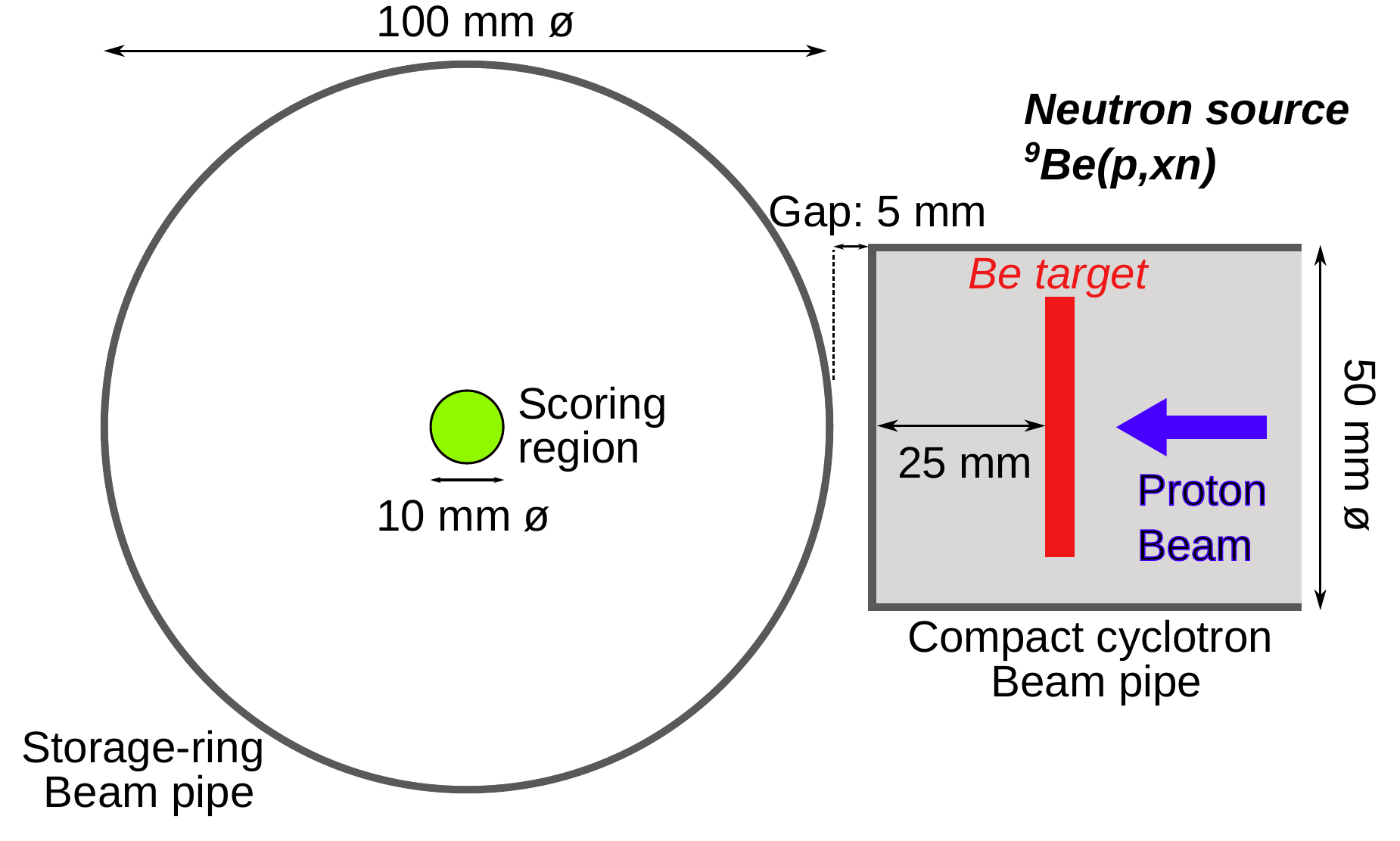}
}
\caption{Cross-sectional schematic showing the storage-ring beam pipe ($\varnothing$ 100\,mm) and the cyclotron beam pipe ($\varnothing$ 50\,mm) with a 5\,mm gap between them. The $^9$Be$(p,xn)$ neutron source is positioned 25\,mm from the cyclotron beam pipe end cap. The central scoring region ($\varnothing$ 10\,mm) indicates the Monte Carlo simulation volume for neutron density calculations.}

    \label{fig:beam_pipe_setup}
\end{figure}

\subsection{\label{sec:RingBeamPipe}  Storage-ring beam pipe design}

The storage-ring beam pipe serves as the primary neutron interaction volume, defining the spatial region where circulating radioactive ions encounter the thermal neutron flux. This subsystem must accommodate the dual requirements of maintaining vacuum conditions for ion beam circulation while maximizing thermal neutron density for nuclear reaction measurements.
The geometrical configuration features a storage-ring beam pipe with a diameter of 100\,mm and a 1\,mm thick aluminum wall, as illustrated in Figure \ref{fig:beam_pipe_setup}. The aluminum wall thickness represents an optimized balance between structural integrity for vacuum containment and minimal neutron absorption. For the design simulations, the storage-ring beam pipe section is assumed to have a length of 200\,cm, providing adequate interaction length for neutron-capture measurements.
Within the storage-ring beam pipe center, a cylindrical void scoring volume of 10\,mm diameter is defined for Monte Carlo calculations to evaluate the thermal neutron density. This scoring volume represents a reasonable computational approximation, considering that the radioactive ion beam circulating in the storage ring is expected to be well-focused with a diameter of a few millimeters.

\subsection{\label{sec:nsource} Compact cyclotron neutron source}
Neutrons are typically produced in-situ through charged-particle induced nuclear reactions. A crucial criterion for selecting the appropriate nuclear reaction is its ability to provide the highest possible neutron yield. Furthermore, it is highly advantageous if the nuclear reaction  generates neutrons with energies as low as possible. This low primary energy facilitates a more effective slowing-down and subsequent thermalization of the neutrons, allows for a more compact moderator design, and significantly reduces neutron leakage from the moderator during the slowing-down process. Consequently, a fundamental physical input for optimizing the moderator geometry involves a trade-off between the neutron yield of the reaction and the primary energy spectrum of the produced neutrons.

The selection of such a nuclear reaction has been extensively studied within the context of developing Compact Accelerator-driven Neutron Sources (CANS)\,\cite{CANS_IAEA} and for applications in Boron Neutron Capture Therapy (BNCT)\,\cite{BNCT_IAEA}. For proton-induced reactions below 30\,MeV, the  ${}^{9}$Be$(p,xn)$ and ${}^{7}$Li$(p,xn)$ reactions stand out as the two most promising neutron-producing reactions due to their high cross sections\,\cite{zakaleketal2020_energy}. Specifically, for very low proton energies (below 5\,MeV), a lithium target is generally preferred and commonly employed for neutron production in various facilities. Conversely, for higher proton energies, up to 30\,MeV, a beryllium target is typically the chosen material at different neutron sources. From the perspective of primary neutron energy, both lithium and beryllium targets yield typically softer neutron energy spectra than fusion or spallation reactions\,\cite{zakaleketal2025_neutron}. Regarding technical realization, the use of beryllium simplifies target design considerably, as this material is generally more stable and easier to handle than lithium. Recent studies have successfully shown the convenience of a three-layer Be-target design\,\cite{Kumada15} to avoid blistering effects arising from the insolubility of H in Be. A similar approach could be pursued here (see Sec.\ref{sec:implementation}).

Over the past 15 years, proton cyclotron technology, particularly for medical isotope production, has undergone significant evolution, trending towards more compact and miniaturized designs. This advancement is largely driven by the high demand for on-site radioisotope production within medical centers. Commercially readily available technology, optimized for producing Positron Emission Tomography (PET) isotopes such as ${}^{18} $F, typically accelerates protons in the range of 7.5 to 12\,MeV, with beam currents spanning from 5\,µA up to 150\,µA\,\cite{oliver2017_compact}. This technology has been specifically designed for compactness and modularity, thereby meeting the stringent installation requirements of hospital environments. Furthermore, recent designs are highly optimized for small footprints and often incorporate self-shielding, significantly enhancing radiation protection and simplifying handling. An interesting example of such compact technology is the recently released IBA Cyclone KEY\,\cite{nuttensetal2023_development}, which features an energy of the extracted protons of 9.2\,MeV and a current of 130\,µA, with a footprint of just $1.5\times 1.4$\,m$^2$. Crucially, the specifications of these low-energy medical cyclotrons align well with our technical requirements in terms of beam energy, current, and footprint, for driving a neutron source based on the ${}^9 $Be$(p,xn)$ reaction. Consequently, our design will focus on assuming proton energies up to 10\,MeV.

The cyclotron beam pipe configuration, as illustrated in Figure\,\ref{fig:beam_pipe_setup}, consists of a 50\,mm diameter tube with 1\,mm thick aluminum walls, positioned perpendicular to the storage-ring beam pipe and centered within the 200\,cm section considered in the simulations. The cyclotron beam pipe maintains a 5\,mm gap between its end cap and the storage-ring beam pipe wall. The beryllium target is housed within the cyclotron beam pipe, positioned 25\,mm from the end cap (see also Sec.\ref{sec:implementation}). 

In this work, for the sake of computational efficiency, the neutron source is modeled as a point source located at the beryllium target center, with energy distributions for the $^9$Be$(p,xn)$ reaction derived from \texttt{Geant4} calculations for low-energy proton bombardment with proton energies $E_p$ in the 5--10\,MeV range\,\cite{luetal2021_geant4}. This simplified--yet physically meaningful--approach enables a systematic evaluation of the moderator performance while maintaining a computational control for extensive parametric studies of the free-neutron target optimization.

\subsection{\label{sec:nModerator} Neutron moderator materials}
The material selection for the moderator geometry optimization represents a critical design consideration that directly influences system performance and operational viability. The moderator must efficiently thermalize fast neutrons while minimizing parasitic absorption to maximize neutron density within the free-neutron target located in the moderator's central region. Moreover, adopting a cost‑efficient strategy is crucial to ensure the feasibility of the demonstrator‑scale prototype.

Effective moderation requires materials exhibiting a low average number of collisions for neutron slowing down, high macroscopic scattering cross-sections ($\Sigma_s$), and correspondingly low macroscopic absorption cross-sections ($\Sigma_a$). These critical neutronic properties are quantitatively evaluated through two figures of merit: the \textit{Macroscopic Slowing Down Power} (MSDP = $\xi\Sigma_s$) and the \textit{Moderating Ratio} (MR = $\xi\Sigma_s/\Sigma_a$)\,\cite{Jevremovic2009}, where $\xi$ represents the average logarithmic energy loss per collision. The MSDP directly quantifies neutron moderation performance, with larger values indicating superior efficiency in energy degradation. The MR provides a comprehensive measure of moderator effectiveness, where high values indicate excellent moderation with minimal parasitic absorption---crucial for maximizing thermal neutron flux. Nuclear properties of some of the potential moderator materials for the free-neutron target are given in Table\,\ref{tab:moderator_properties}.

\begin{table}[htbp]
\centering
\caption{Neutron moderation properties for different materials in the epithermal region and number of collisions ($N_{\text{col}}$) to slow a 2\,MeV neutron down to 1\,eV. Data retrieved from\,\cite{Sharma2001}.}
\label{tab:moderator_properties}

\begin{tabular}{lccc} 
\hline

Material & 
$N_{\text{col}}$ & 
MSDP & 
MR \\
\hline

Light water (\ce{H2O}) & 
16 & 
142.8 & 
62.1 \\

Heavy water (\ce{D2O}) & 
28 & 
17.9 & 
4824.3 \\

Beryllium (Be) & 
70 & 
15.3 & 
127.7 \\

Beryllium oxide (BeO) & 
84 & 
12.2 & 
158.6 \\

Graphite (C) & 
92 & 
8.4 & 
220.4 \\
\hline

\end{tabular}
\end{table}

For the design optimization study of the free-neutron target, the primary materials under investigation include light water (\ce{H2O}), heavy water (\ce{D2O}), beryllium oxide (\ce{BeO}), and graphite (\ce{C}). While metallic beryllium exhibits excellent nuclear properties (see Table\,\ref{tab:moderator_properties}), its extreme toxicity and complex handling requirements\,\cite{stearney2023etal_beryllium} would compromise the feasibility objectives for a demonstrator prototype at storage-ring facilities. Although all beryllium-containing compounds present inhalation hazards, this risk is form-dependent: the OSHA beryllium standard (29~CFR~1910.1024) explicitly exempts articles in non-particulate solid form that are not machined on-site\,\cite{OSHA1910}. In the present design, the \ce{BeO} material is procured as a sintered ceramic component from a qualified commercial supplier, requiring no on-site powder handling or machining. The selection among these candidate materials is carried out through a systematic optimization study presented in Sec.\,\ref{sec:SingleMod}, where neutron moderation performance and practical implementation constraints are evaluated jointly to identify the most suitable configurations.

\subsection{\label{sec:MCsimulations} Monte Carlo simulations}
For the present study, we employed \texttt{ParticleCounter}\,\cite{particlecounter2024}, a specialized \textit{Geant4}-based Monte Carlo simulation tool developed for neutron transport in matter, detector design and optimization, and detection efficiency calculations. Experimentally validated for neutron detector simulations, it achieves precisions within 3\% relative to experimental efficiency measurements and has been successfully applied in projects such as BRIKEN\,\cite{Tarifeno17} and HENSA\,\cite{hensa2024} for conceptual design and geometric optimization of detection systems in relevant nuclear physics energy ranges.  For neutrons up to 20\,MeV, the code utilizes the high-precision G4ParticleHP model with evaluated nuclear data libraries. The simulations reported here used \texttt{ParticleCounter} version~5.4, compiled with \textit{Geant4} version~11.1.3 and the G4NDL4.7 data library, which incorporates thermal scattering data from the JEFF-3.3 library supplemented by the ENDF/B-VIII.0 library for missing materials, applied at the appropriate physical temperature for each moderator material. The accuracy of the \textit{Geant4} thermal neutron transport with these libraries has been validated against reference Monte Carlo codes across the relevant moderator materials and temperature range~\cite{thulliez2022_geant4}. This configuration ensures accurate modeling of all relevant neutron interactions, including elastic and inelastic scattering, capture, and moderation (see Secs.\,\ref{sec:SingleMod}--\ref{sec:CryoMod}).

The primary objective of the Monte Carlo calculations is to determine the density of thermal neutrons, normalized to the primary neutron source strength, within the cylindrical void scoring volume (see Sec.\,\ref{sec:RingBeamPipe}). The cylindrical void scoring volume is subdivided into 100 discrete cells, each 2\,cm in length along the storage-ring beam pipe axis. Within each scoring cell, neutron energy and track length are recorded for all neutrons that intercept the cell volume. The neutron energies define velocity groups for the track-length estimator implementation, while the recorded track-lengths provide the fundamental input for the neutron density estimator normalized to source strength (see definition in Appendix\,\ref{app:density-estimator}). Throughout this work, we define \emph{thermal} neutrons as those with energy $E \leq 1$~eV. The thermal neutron density in the $j$-th scoring cell is denoted as $n_{\mathrm{th},j}$. From this quantity, the corresponding thermal neutron areal density is
\begin{equation}
A_{\mathrm{den},j} = n_{\mathrm{th},j}\, \ell ,
\end{equation}
where $l$ is the length of the scoring cell. The total thermal neutron areal density in the target is then obtained by summing over all scoring cells,
\begin{equation}
A_{\mathrm{den}} = \sum_{j=1}^{100} A_{\mathrm{den},j}.
\end{equation}
The calculation of $A_{\mathrm{den}}$ is fundamental as it constitutes a key ingredient for determining the neutron-induced reaction rate in the storage-ring experiments considered here. The optimization calculations simulate $10^5$ neutron histories per proton energy configuration (5--10~MeV), yielding typical statistical fluctuations in $A_{\mathrm{den}}$ of 1--3\%. Each neutron history is sampled from the beryllium target position. Unless otherwise stated, $A_{\mathrm{den}}$ values are normalized per \emph{primary neutron in the source}, which is indicated in the units as ``/prim''.

The neutron energy distributions are taken from the $^9$Be$(p,xn)$ reaction calculations reported in Ref.\,\cite{luetal2021_geant4} (see Sec.\,\ref{sec:nsource}), where Geant4 simulations employing the \texttt{QGSP\_BIC\_AllHP} physics list were shown to reproduce experimental neutron yields and MCNP6 results within 10\% in the 5--10\,MeV proton energy range. The angular distribution of emitted neutrons is assumed to be uniform in azimuth from $0$ to $2\pi$ around the axis defined by the proton beam direction in the cyclotron beam pipe (see Figure\,\ref{fig:beam_pipe_setup}), as a simplifying approximation. We note that a more comprehensive experimental characterization of the angular and energy distributions of the $^9$Be$(p,xn)$ reaction persists as an unresolved issue, as further discussed in Sec.\,\ref{sec:summary}. 

\subsection{\label{sec:SingleMod} Single-material moderator studies}
\begin{figure}
    \resizebox{0.35\textwidth}{!}{\includegraphics{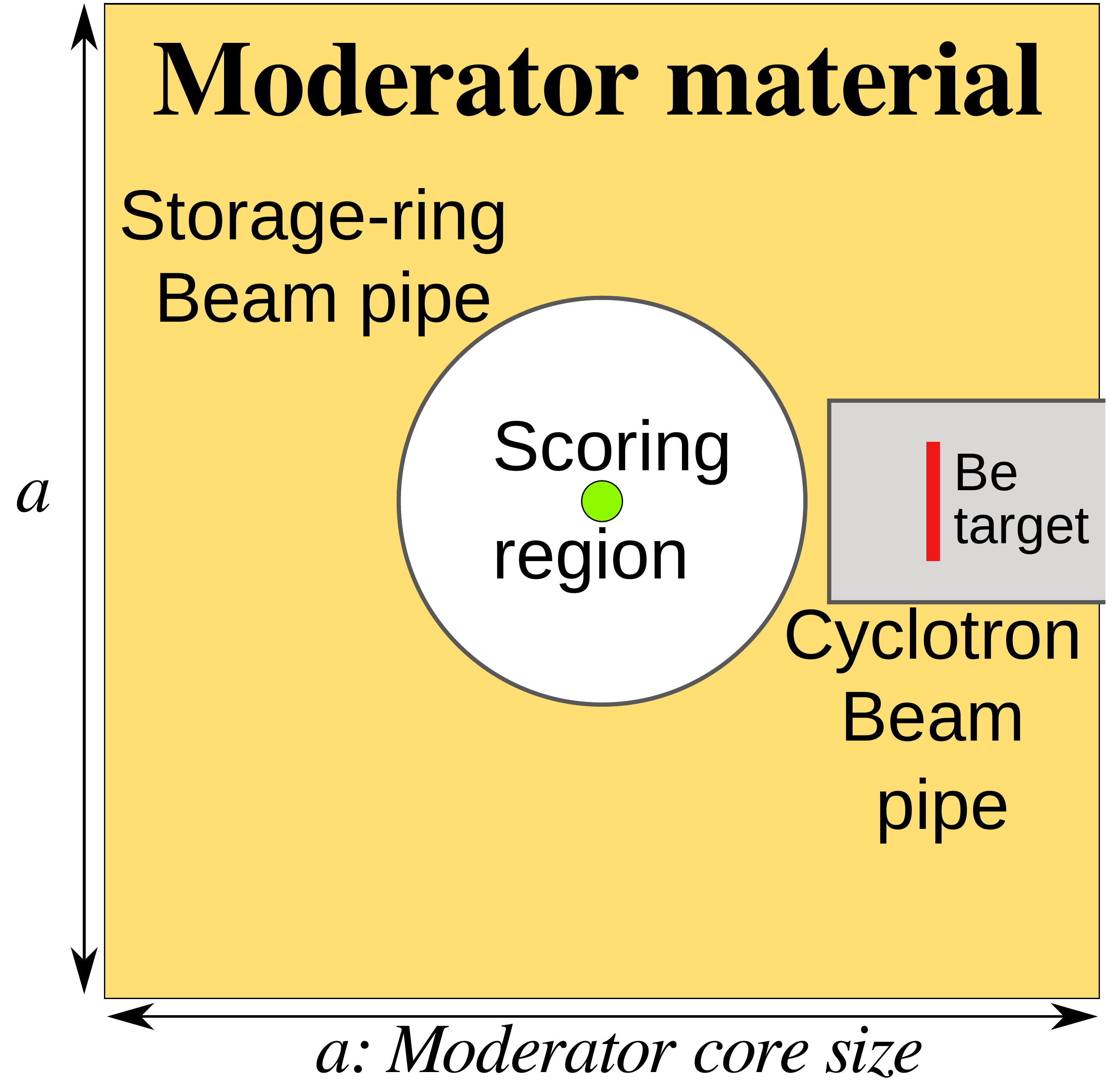}
}
\caption{Cross-sectional schematic of the single-material moderator configuration studies. The moderator consists of a 200\,cm × \(a\) × \(a\) solid parallelepiped containing both beam pipe assemblies, with \(a\) systematically varied to determine optimal dimensions for each material.}
    \label{fig:SingleMod_setup}
\end{figure}
\begin{figure}
    \resizebox{0.5\textwidth}{!}{\includegraphics{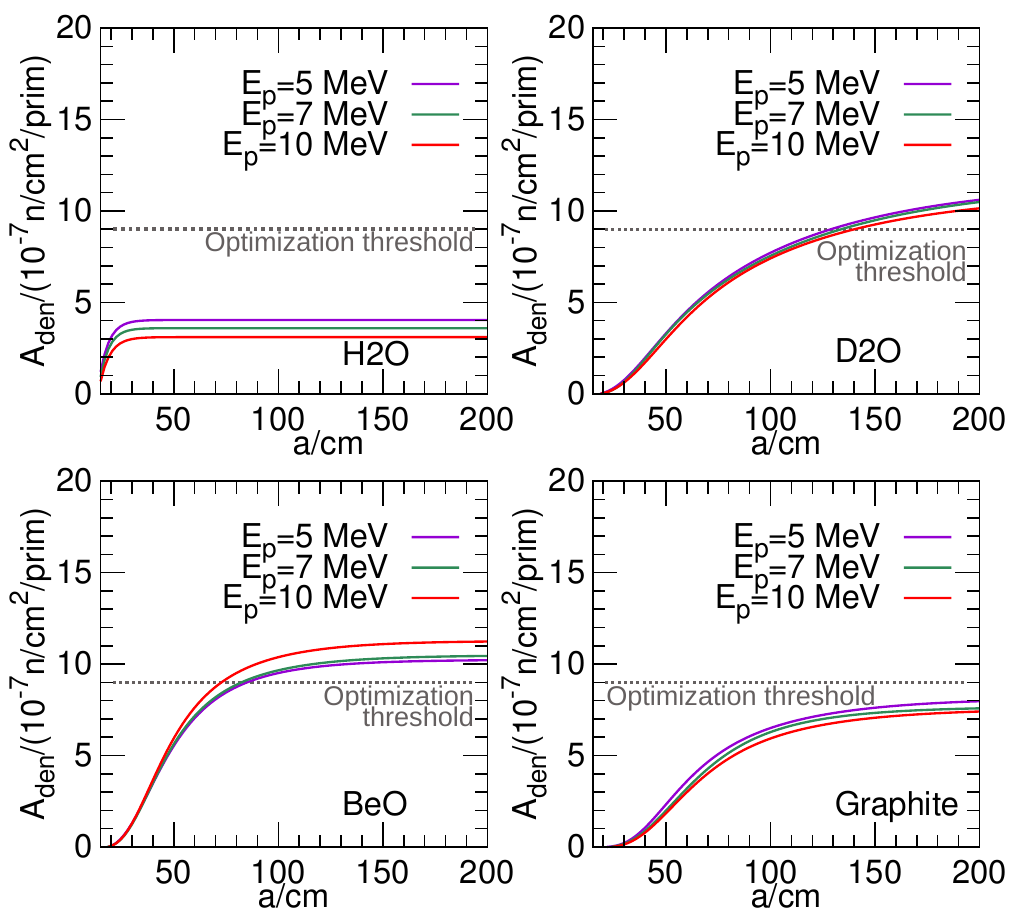}
}
\caption{Monte Carlo calculation results for single-material moderator design. The dotted black line represents the threshold areal density for a compact and cost-effective moderator.}
    \label{fig:MCresults_singleMod}
\end{figure}
The first optimization stage establishes asymptotic performance limits for individual moderator materials and identifies suitable candidates for both moderator and reflector roles in compact, cost-effective designs.

The single-material studies employ the standardized geometry illustrated in Figure\,\ref{fig:SingleMod_setup}, featuring a 200\,cm $\times$ \textit{a} $\times$ \textit{a} solid parallelepiped moderator containing both beam pipe and cyclotron pipe assemblies. The thermal areal density in the central void volume has been systematically studied for \ce{H2O}, \ce{D2O}, \ce{BeO}, and graphite. The moderator core size parameter (\textit{a}) was varied across all materials in the range spanning from 18\,cm to 200\,cm to establish comprehensive performance characterization while maintaining the fixed 200\,cm length corresponding to the storage-ring beam pipe interaction region.

The Monte Carlo results, shown in Figure\,\ref{fig:MCresults_singleMod}, reveal that each moderator material has an asymptotic performance limit that depends only slightly on proton energy but strongly on the material itself. Here, performance refers to the thermal neutron areal density $A_{\mathrm{den}}$, and the asymptotic value corresponds to the saturation level reached in the parametric scans for increasing material dimensions. Mathematical analysis of the Monte Carlo results yields the following asymptotic limits: \ce{D2O} ($11.6 \times 10^{-7}$\,n/cm$^2$/prim), \ce{BeO} ($10.2$--$11.3 \times 10^{-7}$\,n/cm$^2$/prim), graphite ($7.6$--$8.2 \times 10^{-7}$\,n/cm$^2$/prim), and \ce{H2O} ($3.1$--$4.0 \times 10^{-7}$\,n/cm$^2$/prim). 

The Monte Carlo simulations exhibit an energy dependence trend--- lower proton energy yields higher thermal areal density ---for all materials except \ce{BeO}. This trend reflects neutron spectrum hardening at higher proton energies, which increases neutron diffusion and leakage from the moderator. \ce{BeO} deviates from this pattern due to the effect of neutron multiplication reactions.

The simulation results for the different materials can be summarized as follows:

\begin{itemize}
    \item \textbf{Light water (\ce{H2O})}. Saturates at just 40\,cm but remains well below the optimization threshold due to intrinsic neutron absorption. Excluded as a primary moderator candidate.

    \item \textbf{Heavy water (\ce{D2O})}. Consistently exceeds the optimization threshold across all proton energies, reaching 80\% of its asymptotic performance at an average dimension of 138\,cm.

    \item \textbf{Beryllium oxide (\ce{BeO})}. Displays an inverse proton energy trend due to neutron multiplication via $\mathrm{Be}(n,2n)$ reactions, most apparent at 10\,MeV proton energy. Reaches 80\% of asymptotic performance at 72\,cm, making it especially attractive for compact designs.

    \item \textbf{Graphite (C)}. Does not reach the optimization threshold (80\% of asymptotic at $a=100$\,cm on average), but its low cost and favorable neutron transport properties make it an excellent reflector candidate.
\end{itemize}

For compact, cost-effective designs optimized for storage-ring integration, an asymptotic optimization threshold of 9 $\times$ 10$^{-7}$\,n/cm$^2$/prim is established. This threshold represents a practical performance level achievable in compact configurations while maintaining feasibility for storage-ring integration. The selected value corresponds to approximately 80\% of the asymptotic performance for both \ce{D2O} and \ce{BeO}, ensuring that primary moderator candidates can achieve efficient operation within the dimensional constraints imposed by storage-ring facility requirements.

\subsection{\label{sec:ModRef} Moderator/reflector combination studies}
\begin{figure}
    \resizebox{0.5\textwidth}{!}{\includegraphics{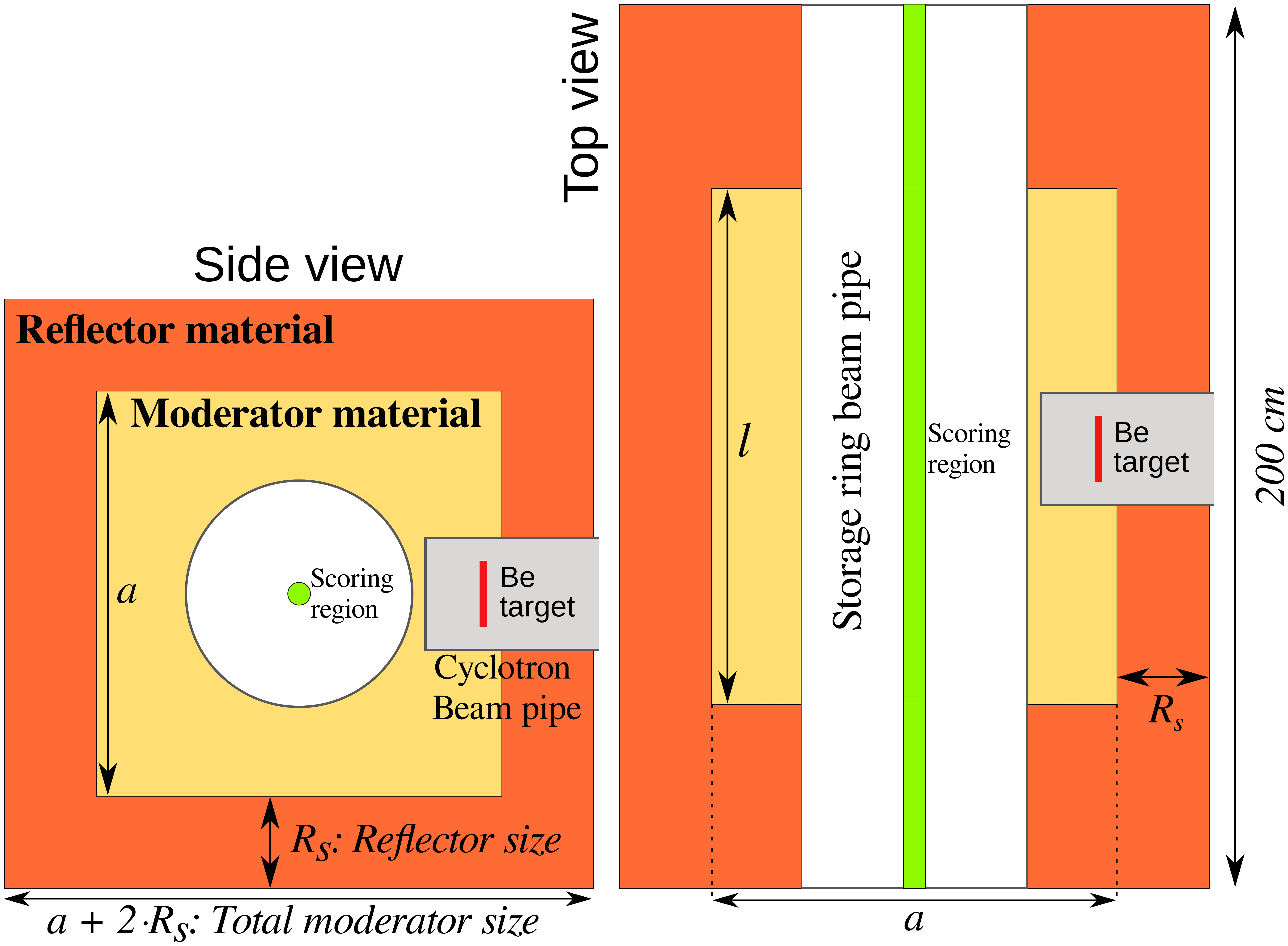}
}
\caption{Cross-sectional schematic for the moderator-reflector configuration studies. The neutron target assembly consists of a 200\,cm × (\(a+2R_s\)) × (\(a+2R_s\)) solid parallelepiped containing both beam pipe assemblies, with \(a\), \(l\), and \(R_s\) systematically varied to determine optimal dimensions for each material.}
    \label{fig:ModRef_setup}
\end{figure}
\begin{figure}
    \resizebox{0.4\textwidth}{!}{\includegraphics{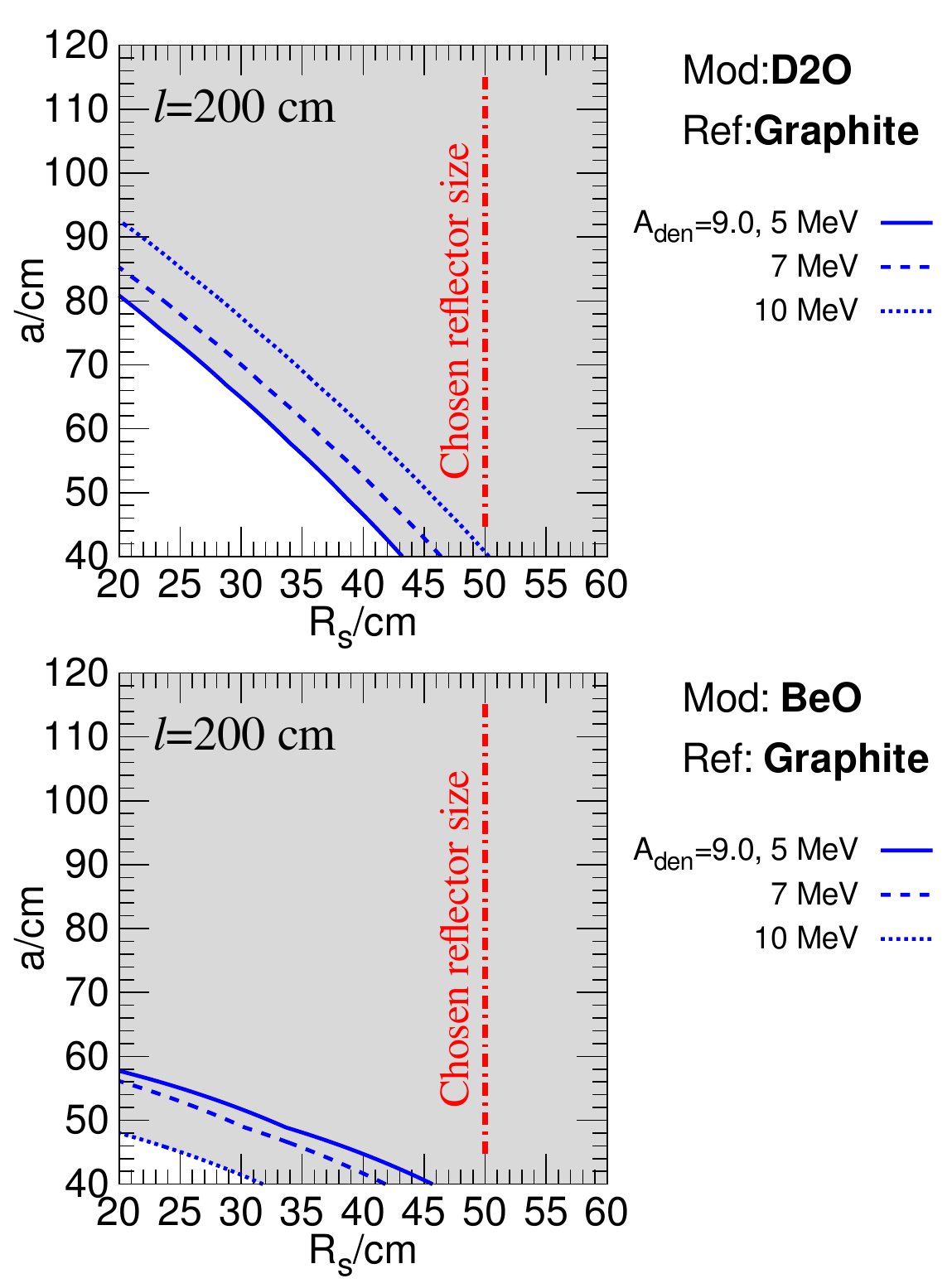}
}
\caption{Monte Carlo optimization results for moderator/reflector combinations keeping $l=200$~cm. Blue contours represent the asymptotic optimization threshold of $9 \times 10^{-7}$\,n/cm$^2$/prim for 5, 7, and 10\,MeV proton energies. Gray shaded regions show configurations exceeding the threshold for \ce{D2O} + graphite and \ce{BeO} + graphite combinations across the parameter space $a = 40–120$\,cm and $R_s = 20–60$\,cm.}
    \label{fig:MCresults_ModRef_conf100}
\end{figure}

\begin{figure}
    \resizebox{0.4\textwidth}{!}{\includegraphics{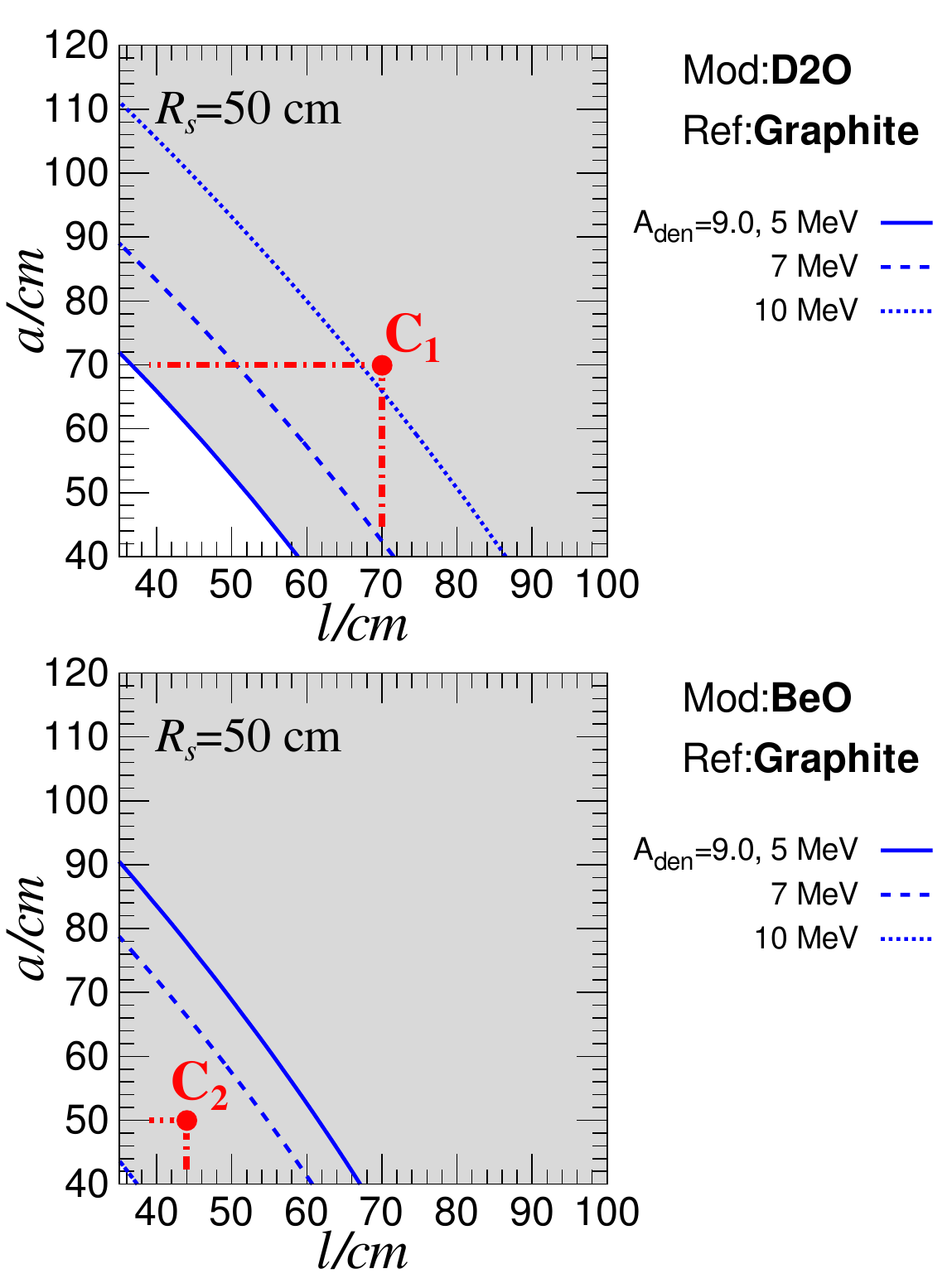}
}
\caption{Monte Carlo calculation results for moderator materials \ce{D2O} and \ce{BeO} and graphite reflector using chosen reflector size $R_s=50$\,cm. Selected candidate configurations, displayed in red color, are $C_1$ and $C_2$, corresponding to core moderators of  \ce{D2O} and \ce{BeO}, respectively. The contour lines, in blue, corresponds to $A_{\mathrm{den}}=9.0$ in units of $10^{-7}$~n/cm$^{2}$/prim for different proton energies.}
    \label{fig:MCresults_ModRef_conf101}
\end{figure}
The second optimization stage investigates moderator/reflector combinations to enhance thermal neutron density beyond the single-material asymptotic limits established in Sec.\,\ref{sec:SingleMod}, while maintaining compact geometries 
suitable for storage-ring integration.

Based on the results from Sec.\,\ref{sec:SingleMod}, two moderator/reflector combinations are selected for comprehensive study: (1) \ce{D2O} moderator + graphite reflector, combining high-performance moderation with cost-effective reflection; and (2) \ce{BeO} moderator + graphite reflector, exploiting compact geometry with neutron multiplication enhancement potential. These configurations leverage the established moderator candidates that exceed the asymptotic optimization threshold while incorporating graphite as a reflector material due to its favorable neutron transport properties, as demonstrated in the single-material analysis, and its inherent cost-effectiveness.

The moderator/reflector studies employ a systematic two-phase optimization approach using the geometry illustrated in Figure \ref{fig:ModRef_setup}, featuring a 200~cm $\times$ $(a + 2R_s)$ $\times$ $(a + 2R_s)$ rectangular block assembly containing both beam pipe assemblies. The configuration consists of a central moderator core of dimension $a \times a \times l$ surrounded by a reflector shell of thickness $R_s$, creating the total system dimensions. The first phase investigates the thermal areal density ($A_{den}$) across the parameter space $a = 40$--120~cm and $R_s = 20$--60~cm, maintaining $l = 200$~cm constant. This configuration exploits the excellent neutron transport properties of the moderator materials, although it may result in less cost-efficient configurations due to the large volumes of \ce{D2O} or \ce{BeO} required.

The Monte Carlo results for the first phase, presented in Figure \ref{fig:MCresults_ModRef_conf100}, demonstrate the optimization landscape for both moderator/reflector combinations across the investigated parameter space. The figure includes contour curves (shown in blue) representing the asymptotic optimization threshold of $9 \times 10^{-7}$~n/cm$^2$/prim for each proton energy, while the gray shaded region indicates the configuration space achieving performance above the optimization threshold. The systematic analysis reveals that a reflector thickness of $R_s = 50$~cm maximizes the areal density as a function of parameter $a$, exceeding the optimization threshold for all proton energies in the 5--10~MeV range. Based on these results, $R_s = 50$~cm is selected as the optimal reflector thickness for the second phase studies.

The second phase focuses on optimizing both moderator core dimensions to maximize areal density while minimizing the total moderator volume for cost-effective implementation. With the optimal reflector thickness of $R_s = 50$~cm established in the first optimization phase, the parameter space is explored across $a = 40$--120~cm and $l = 35$--100~cm. The Monte Carlo results for this optimization study are presented in Figure \ref{fig:MCresults_ModRef_conf101}, employing an analogous representation to Figure \ref{fig:MCresults_ModRef_conf100} with blue contour curves indicating the asymptotic optimization threshold and gray shaded regions showing viable configuration spaces for both moderator/reflector combinations. 

From this analysis, two candidate configurations are selected for further optimization. Configuration $C_1$, shown in the top panel of Figure \ref{fig:MCresults_ModRef_conf101}, represents the selected candidate for the \ce{D2O} moderator combination. Based on the calculation results, $C_1$ ($a = 70$~cm, $l = 70$~cm) provides an optimal trade-off between areal density exceeding the optimization threshold and relatively reduced \ce{D2O} volume requirements. Configuration $C_2$, presented in the bottom panel of Figure \ref{fig:MCresults_ModRef_conf101}, corresponds to the \ce{BeO} moderator and requires different selection criteria. Given the neutron multiplication capabilities of beryllium at proton energies above 7~MeV, viable configurations span practically the entire studied parameter space for these higher energies. Therefore, $C_2$ selection prioritizes a minimum \ce{BeO} volume while maintaining compatibility with proton energies above 7~MeV. The analysis concludes that $C_2$ ($a = 50$~cm, $l = 44$~cm) represents an optimal balance between areal density above the threshold, reduced \ce{BeO} volume, and performance optimization for proton energies above 7~MeV. These candidate configurations establish the foundation for the third optimization stage incorporating cryogenic enhancement to further maximize thermal neutron density.

\subsection{\label{sec:CryoMod} Cryogenic moderator enhancement study}
\begin{figure}
    \resizebox{0.3\textwidth}{!}{\includegraphics{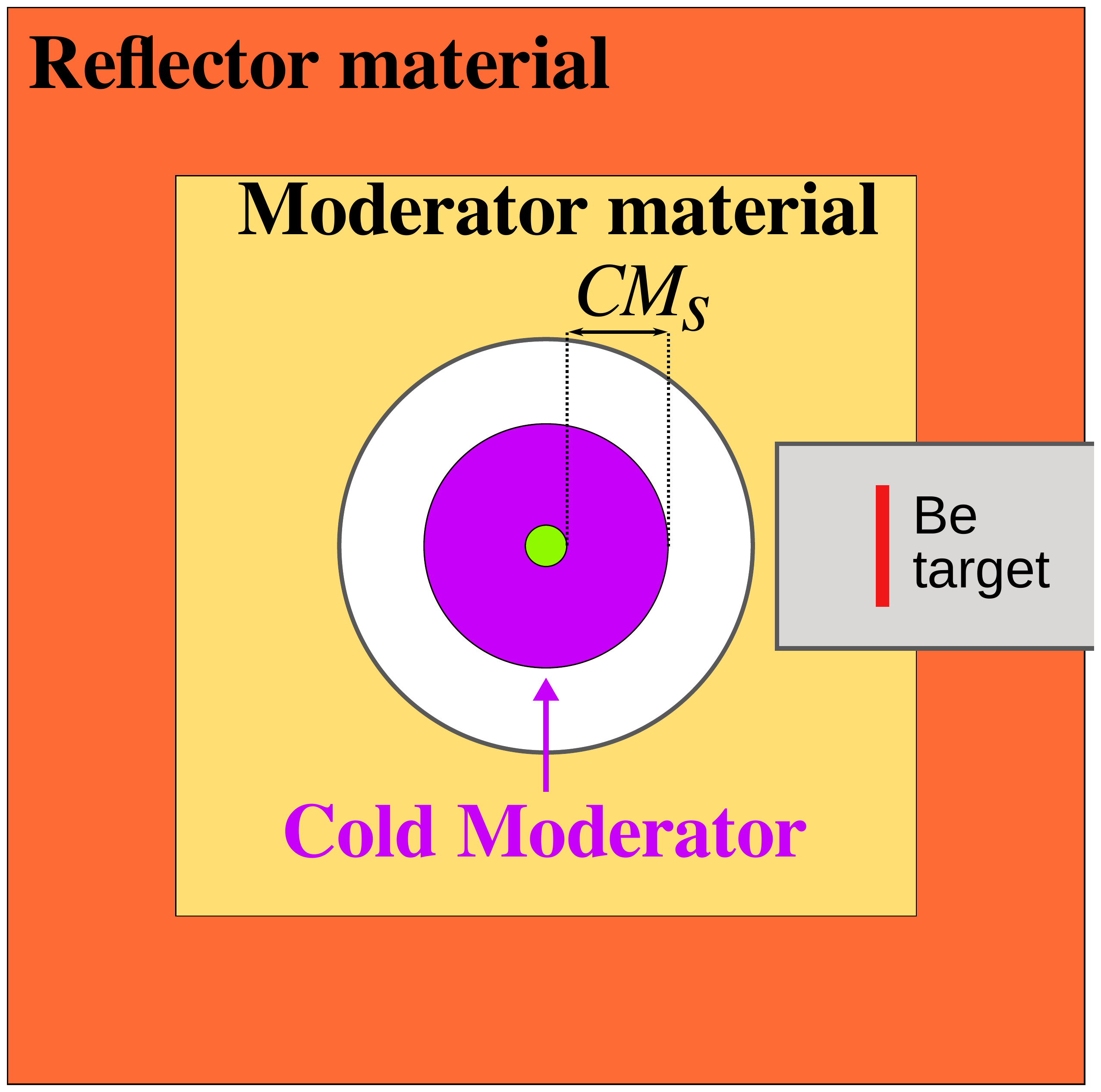}
}
\caption{Cross-sectional schematic of the moderator configuration studies. The moderator consists of a 200 cm × \(a\) × \(a\) solid parallelepiped containing both beam pipe assemblies, with \(CM_s\) systematically varied to determine optimal dimensions for each material.}
    \label{fig:Setup_ModRefconf_cryo}
\end{figure}

The third optimization stage incorporates a cryogenic \ce{LH2} moderator to further maximize thermal neutron density beyond the performance of the optimized moderator/reflector combinations established in Sec.\,\ref{sec:ModRef}.

Liquid hydrogen (LH$_2$) at 20 K represents the optimal choice for cold neutron moderators due to its exceptional hydrogen nucleus density and unique scattering behavior. At this temperature, hydrogen exists predominantly as para-hydrogen ($>$99\%), which exhibits filter-like behavior with at least a 30-fold reduction in scattering cross-section below standard room thermal energy range. This enables the construction of \textit{low-dimensional} moderators where cold neutrons can easily travel through the moderator material with minimal risk of regaining energy through multiple collisions \cite{CANS_IAEA}. The large hydrogen scattering cross-section ensures superior moderation effectiveness, while LH$_2$ offers greater radiation resistance and operational safety advantages over alternative cold moderator materials such as solid methane.

The cryogenic enhancement study employs the candidate configurations $C_1$ and $C_2$ established from the moderator/reflector optimization as the baseline geometries for systematic investigation, see Sec.\,\ref{sec:ModRef}. The cold moderator features a cylindrical geometry that is completely embedded within the storage-ring beam pipe and directly surrounds the void scoring volume through which the radioactive ion beam circulates, as illustrated in Figure~\ref{fig:Setup_ModRefconf_cryo}. This configuration ensures optimal neutron-ion interaction while providing a vacuum layer which is essential for thermal isolation of the cold moderator. The cold moderator thickness (CM$_s$) is varied over 0--40\,mm, with para-hydrogen thermalization at 20\,K treated using the thermal scattering libraries described in 
Sec.\,\ref{sec:MCsimulations}.

\begin{figure}
    \resizebox{0.35\textwidth}{!}{\includegraphics{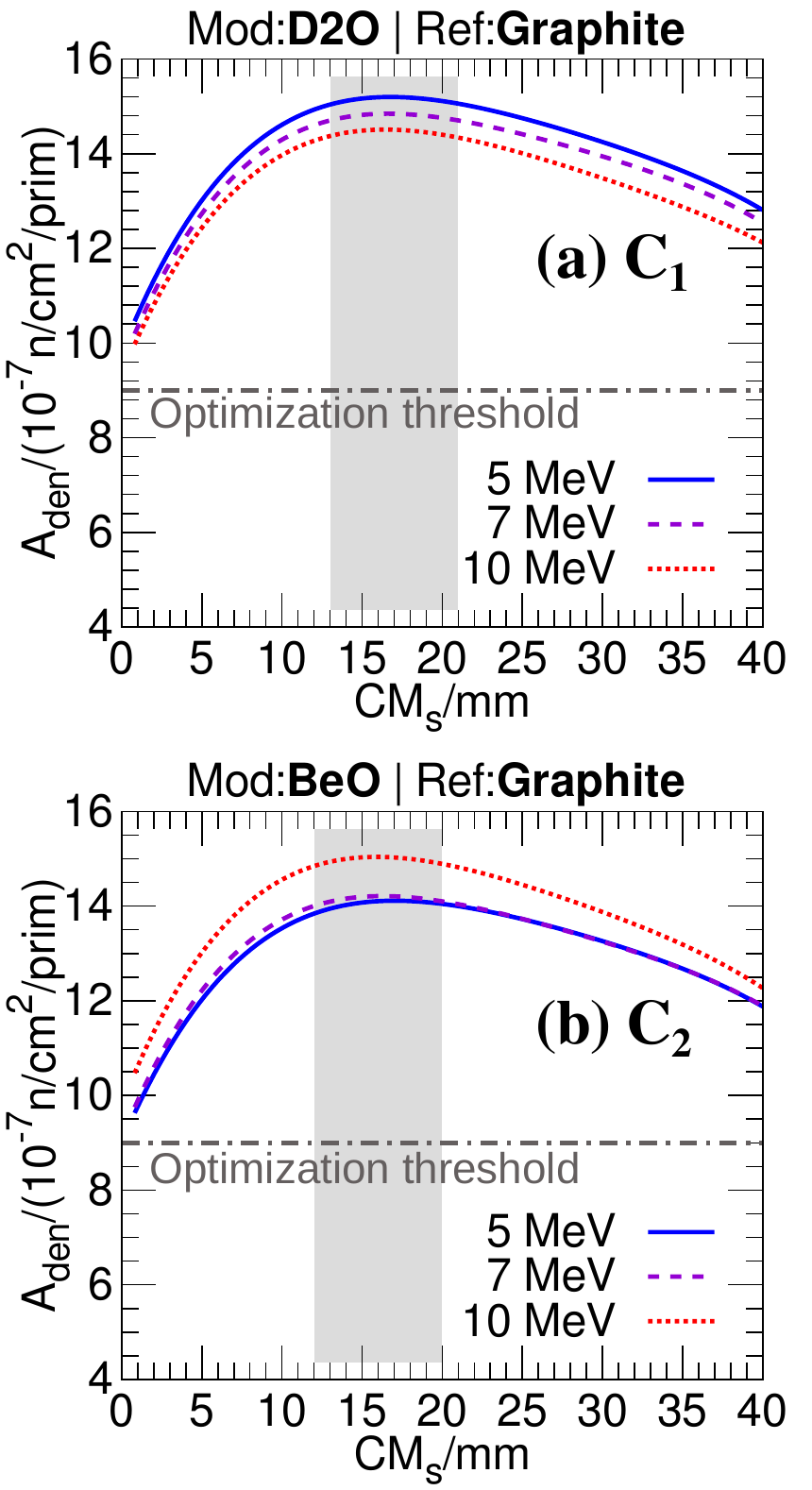}
}
\caption{Monte Carlo calculation results for the candidate configuration, $C_1$ and $C_2$, including a cryogenic \ce{LH2} moderator at 20 $K$. The thickness of the cold moderator ($CM_s$) allows to maximize the thermal areal density. The dotted black line represents the threshold areal density for a compact and cost-effective moderator.}
    \label{fig:MCresults_ModRefconf_cryo}
\end{figure}

The Monte Carlo calculation results for the cryogenic enhancement optimization are presented in Figure~\ref{fig:MCresults_ModRefconf_cryo}. The systematic analysis reveals significant performance improvements ($\sim$60\% over the optimization threshold) for both candidate configurations across the investigated proton energy range. The optimization results demonstrate distinct plateau regions where a maximum areal density is maintained across the range of cold moderator thicknesses. Configuration $C_1$ exhibits optimal performance across the range CM$_s$ = 13--21\,mm, while configuration $C_2$ achieves peak areal density within CM$_s$ = 12--20\,mm, as indicated by the gray shaded regions in Figure~\ref{fig:MCresults_ModRefconf_cryo}. Beyond these plateau regions, additional LH$_2$ thickness provides diminishing returns due to neutron absorption effects. The energy dependence for a \ce{D2O} moderator observed in previous optimization stages persists, generally yielding superior thermal areal densities for lower proton energies. As for the configuration $C_2$, the inverse energy trend characteristic of BeO neutron multiplication effects remains, which is particularly pronounced at 10\,MeV proton energy.

Based on the calculation results, CM$_s$ = 13\,mm is selected as the optimal cold moderator thickness. This value represents an optimized compromise that maximizes thermal areal density for both configurations while minimizing the LH$_2$ volume that must be integrated within the beam pipe assembly, thereby reducing cryogenic system complexity and operational requirements.

\subsection{\label{sec:FinalConfigs} Final configuration of the neutron target}

The three-stage optimization methodology yields two recommended configurations for the free-neutron target, summarized in Table~\ref{tab:final_configs}.

\begin{table*}[htbp]
\centering
\caption{Final proposed configurations for the compact neutron target optimized for storage-ring integration.}
\label{tab:final_configs}
\resizebox{\textwidth}{!}{%
\begin{tabular}{cccccccccccc}
\hline
& \multicolumn{3}{c}{\textbf{Moderator}} & \multicolumn{2}{c}{\textbf{Reflector}} & \multicolumn{3}{c}{\textbf{Cold Moderator}} & \textbf{Proton Energy} & \textbf{Thermal Areal Density} \\
\textbf{Configuration} & Material & $a$ & $l$ & Material & $R_s$ & Material & Temp. & $CM_s$ & $E_p$ & $A_{den}$ \\
& & [cm] & [cm] & & [cm] & & [K] & [mm] & [MeV] & [10$^{-7}$ n/cm$^2$/prim] \\
\hline
$C_1$ & \ce{D2O} & 70 & 70 & graphite & 50 & \ce{LH2} & 20 & 13 & 5--10 & 14.3--15.0 \\
$C_2$ & \ce{BeO} & 50 & 44 & graphite & 50 & \ce{LH2} & 20 & 13 & $>$7 & 14.0--15.0 \\
\hline
\end{tabular}%
}
\end{table*}

\begin{figure}[htbp]
\centering
\includegraphics[width=0.35\textwidth]{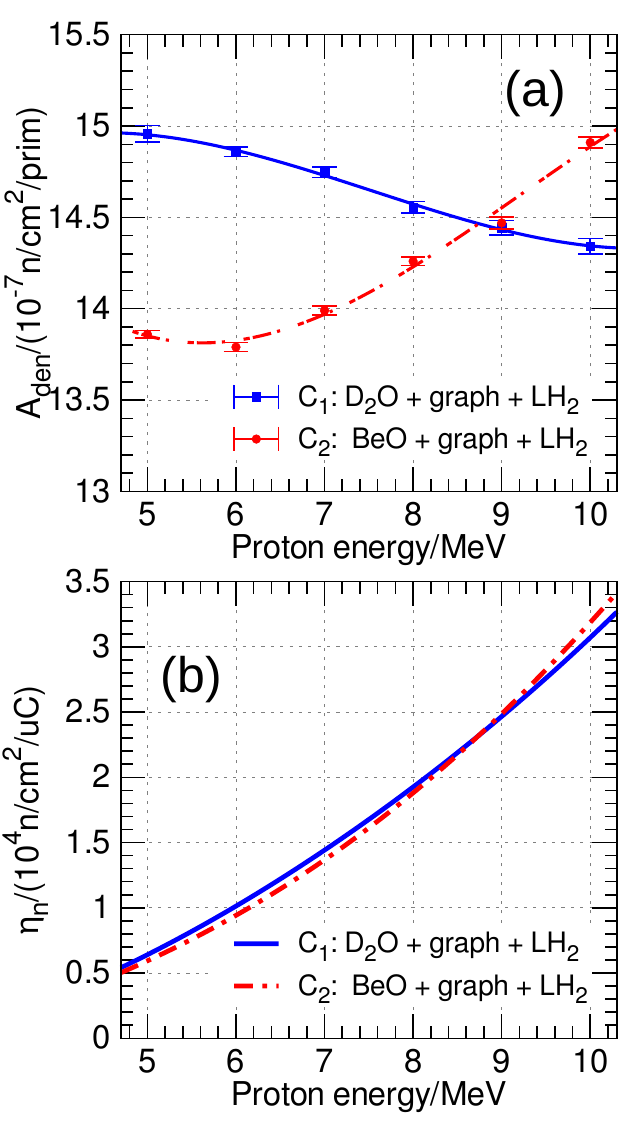}
\caption{(a) Thermal areal density normalized to primary neutron source strength for configurations $C_1$ and $C_2$ as a function of proton energy. Monte Carlo calculations with 4$\times$10$^6$ neutron histories per energy point. (b) Thermal areal density for both configurations folded with the $^9$Be$(p,xn)$ production yields from Ref. \cite{luetal2021_geant4}.}
\label{fig:Results_high_precision}
\end{figure}

Both configurations substantially exceed the optimization threshold, achieving $\sim$60\% above this baseline. High-statistics Monte Carlo calculations (4$\times$10$^6$ neutron histories per energy point, statistical uncertainties below 0.5\%) across the full proton energy range are presented in Figure\,\ref{fig:Results_high_precision}, where the smooth energy dependence of $A_{\mathrm{den}}$ enables reliable polynomial interpolation at intermediate energies.

A market survey among international suppliers of nuclear-grade graphite, \ce{D2O} with purity $>$99.8\%, and solid fired \ce{BeO} ceramic blocks reveals cost ratios per unit volume of approximately 1:100:1000, respectively. Given the effective moderator volume optimization achieved in the proposed configurations, the implementation cost of configuration $C_2$ is approximately a factor of 3 that of configuration $C_1$, as the total system cost is dominated by the moderator material rather than the reflector components.

The optimization process demonstrates that \ce{BeO} does not provide direct advantages under cost-constrained feasibility criteria, as reflected in the similar performance of configurations $C_1$ and $C_2$. However, our calculations indicate that an enhanced areal density can be achieved using larger \ce{BeO} moderator geometries when cost constraints are relaxed. For instance, employing \ce{BeO} with the $C_1$ geometry ($a=70$~cm and $l=70$~cm) yields 16$\times$10$^{-7}$~n/cm$^2$/prim at 10~MeV proton energy, albeit at approximately nine times the cost of configuration $C_1$.

Configuration $C_1$ represents the most cost-efficient option for proton beam energies in the range 5--10~MeV. For applications where budget constraints are not limiting factors, \ce{BeO}-based configurations can deliver superior neutron density performance, though at substantially higher implementation costs. The final selection should prioritize feasibility requirements that balance performance targets against practical budget and operational constraints.

The conversion of these normalized areal densities to absolute target performance is presented in Sec.\,\ref{sec:implementation}, where commercial cyclotron models covering a wide range of beam currents are evaluated to establish the achievable performance for both the proof-of-concept demonstrator and future higher-intensity implementations.

\section{\label{sec:implementation} Implementation of the supercompact-cyclotron driven neutron target}

The transition from conceptual design to operational hardware requires selecting commercially available components and engineering solutions that guarantee reliable operation in a storage-ring environment. The neutron-target concept proposed here has not yet been demonstrated experimentally; its implementation must therefore focus on a proof-of-concept demonstrator that can be integrated into existing facilities with minimal modifications, prioritizing feasibility over ultimate performance.

The implementation approach prioritizes feasibility and modularity, enabling deployment at various storage-ring facilities worldwide while maintaining the performance characteristics established through the Monte Carlo optimization studies presented in Section~\ref{sec:conc_des}. The proposed system leverages mature medical cyclotron technology, proven moderator materials, and established cryogenic systems to deliver a robust, cost-effective neutron target suitable for pioneering neutron-capture measurements in inverse kinematics.

The selection of an appropriate cyclotron represents the most critical component choice for the neutron target implementation.
A survey of commercial compact cyclotrons identifies a clear separation between low-energy systems ($E_p \leq 10$~MeV) intended for medical isotope production and higher-energy research-grade machines. For a first demonstrator, low-energy units offer decisive advantages: compact footprint, low infrastructure requirements, and well-established reliability. Table~\ref{tab:commercial_cyclotrons} summarizes representative commercial models and their estimated thermal areal densities, calculated for $E_p \leq 10$~MeV using the calculation results of Figure\,\ref{fig:Results_high_precision}, and for $E_p > 10$~MeV using a constant per-primary value of $A_{\mathrm{den}}=5\times10^{-7}$~n/cm$^2$/prim combined with neutron yields from Ref.~\cite{Lavelle08}.

\begin{table*}[htbp]
\centering
\caption{Selection of commercial compact proton cyclotrons. Technical specifications and suitability for implementation of the free neutron target are based on proposed configuration with a \ce{D2O} moderator (see configuration $C_1$ Table\,\ref{tab:final_configs}). For proton energies up to 10\,MeV, the total areal density is taken from calculations in Figure\,\ref{fig:Results_high_precision}. For proton energies larger than 10\,MeV, calculations are assuming a thermal areal density per primary neutron $A_{den}=5 \times 10^{-7}$\,n/cm$^2$/prim and neutron yields from Ref.\,\cite{Lavelle08} ($E_p>$10\,MeV).}
\label{tab:commercial_cyclotrons}
\resizebox{\textwidth}{!}{%
\begin{tabular}{lllccclcc}
\hline
\textbf{Manufacturer} & \textbf{Model} & \textbf{Footprint} & \textbf{Weight} & \textbf{Proton Energy} & \textbf{Max Current} & \textbf{Shielding} & \textbf{Thermal Areal Density} & \textbf{Source} \\
 & & ($m^2$) & (tons) & (MeV) & ($\mu$A) &  & (n/cm$^2$) &   \\
\hline
IBA & Cyclone KEY & 1.5x1.4 & 7.5 & 9.2 & 130 & Self-shielding (opt) & $3.4 \times 10^{6}$ & \cite{iba_1_1}  \\
Best & BG-95 & $<$~2x2 & 22 & 9.5 & 120 & Self-shielding & $3.3 \times 10^{6}$ & \cite{best_1_2,teambest_2_2} \\
Best & B6-15/B15p & 2.2x2.2 & 14 & 10 & 400 (550~\footnote{Measured at Argonne National Laboratory}) & Self-shielding  & $1.2 \times 10^{7}$ & \cite{best_1_3}  \\
\hline
GE & PETtrace 800 series & 1.33x1.2  & 20 & 16.5 & 160 & Vault & $6.7 \times 10^{6}$ & \cite{ge_1_4}   \\
IBA & Cyclone KIUBE & 1.9x1.9  & 18 & 18 & 300 & Vault & $1.5 \times 10^{7}$& \cite{iba_1_5}   \\
& &  & & & & Self-shielding (opt) &  &   \\
ACSI & TR-FLEX & 1.7x1.7 & 24 & 22 & 800 & Vault & $6.4 \times 10^{7}$ & \cite{acsi_1_6}   \\
IBA & Cyclone IKON & 2.2x2.2 & 30 & 22 & 1500 & Vault & $1.2 \times 10^{8}$ & \cite{iba_1_7}   \\
ACSI & TR-30 & 2.4x2.4 & 50 & 22 & 1600 & Vault & $1.3 \times 10^{8}$ & \cite{acsi_1_8}   \\
\hline
IBA & Cyclone 70 & 4x4 & 120 & 30-70 & 750 & Vault & -- & \cite{iba_1_9, kek_2_9} \\
\hline

\end{tabular}
}
\end{table*}

\begin{figure*}[!htb]
    \resizebox{0.85\textwidth}{!}{\includegraphics{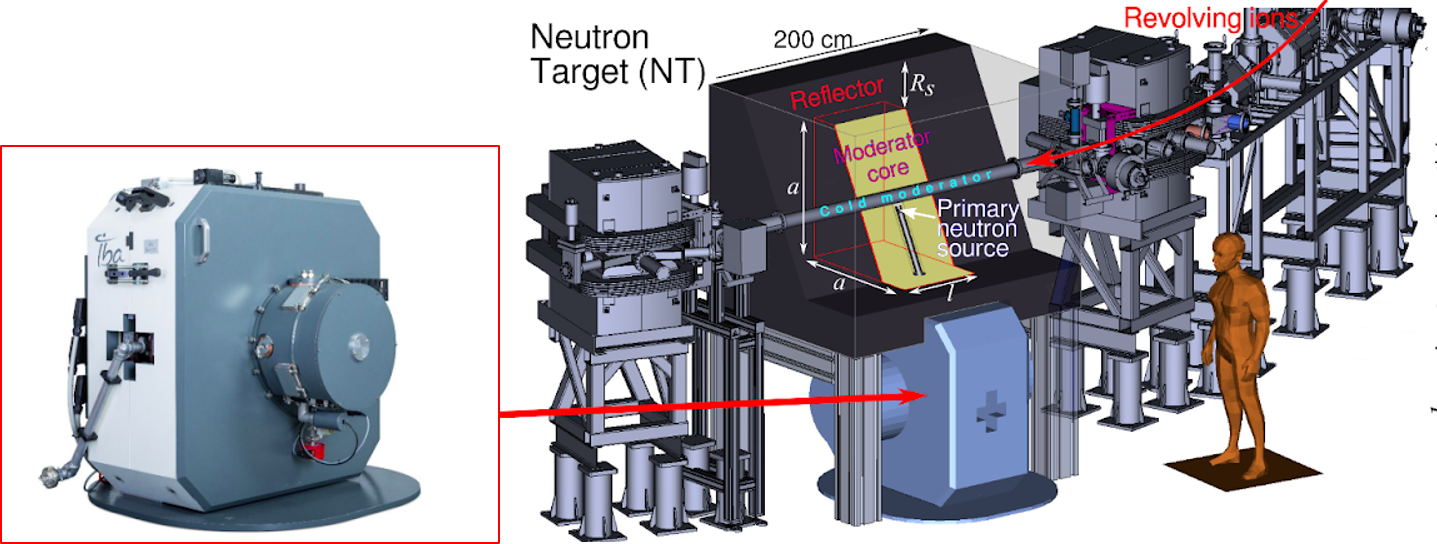}
}
\caption{(Left) The super-compact cyclotron IBA Cyclone KEY \cite{iba22} with a footprint of $\sim$ 2\,m$^2$ and a weight of 7.5\,t. (Right) Sketch of a possible implementation of the full system at the CRYRING ion-storage ring at GSI. For details about the neutron target geometry see Table \ref{tab:final_configs}.}
\label{fig:nstar}       
\end{figure*}

Among the low-energy cyclotron options, the IBA Cyclone KEY emerges as the preferred choice for the proof-of-concept demonstrator due to its exceptional compactness (1.5$\times$1.4\,m$^2$ footprint, 7.5\,t weight), adequate performance (9.2\,MeV protons, 130\,$\mu$A maximum current), and proven reliability in medical applications. The system achieves a thermal areal density of 3.4$\times 10^{6}$\,n/cm$^2$ when operating at maximum current. For operational planning and performance estimates throughout this work, we adopt a conservative nominal value of 3$\times 10^{6}$\,n/cm$^2$, providing margin for system reliability and beam optimization. The compact cyclotron dimensions allow it to be embedded directly within the graphite reflector assembly of configuration $C_1$ (\ce{D2O} moderator, graphite reflector, and \ce{LH2} cold moderator at 20\,K). Figure\,\ref{fig:nstar} shows a possible implementation of the neutron target proposed here. This integrated assembly occupies roughly 2\,m $\times$ 2\,m $\times$ 2\,m, fitting within a typical storage-ring experiment section without major infrastructure changes. The self-contained neutron source is optimally positioned with respect to the storage-ring beam pipe, minimizing neutron transport losses and simplifying alignment.  The small footprint  of the cyclotron makes it particularly well suited for a demonstrator system, thus eliminating the need of a large ancillary facility, as would be required for a spallation source \cite{Reifarth17} or a fission reactor \cite{Reifarth14}.  Finally, it is worth emphasizing that thick beryllium targets are known to suffer from severe hydrogen embrittlement problems~\cite{Rinckel2012} arising from the insolubility of hydrogen in beryllium. This issue can be effectively mitigated through a three-layer target design~\cite{Kumada15}, where nearly all incident protons are stopped in a second layer composed of a suitable material such as palladium. Following this approach, the beryllium target layer is designed to have a thickness corresponding to 87\% of the proton range in beryllium. This configuration preserves a high neutron yield from the $^9$Be$(p,xn)$ reaction while allowing $>99\%$ of the protons to traverse the beryllium layer and deposit their remaining energy in the palladium stopping layer. For 9.2 MeV protons, the projected range in beryllium calculated with SRIM-2013 \cite{SRIM2013} is 688 $\mu$m. The beryllium layer thickness of 0.6 mm was optimized such that the Bragg peak falls entirely within the palladium stopping layer, thereby ensuring that both the maximum energy deposition and the hydrogen implantation peak occur outside the beryllium layer. At this conceptual design stage, detailed modeling of the proton-stopping palladium layer and the copper cooling backing has not been included, although their technical implementation follows established procedures as described in Ref.~\cite{Kumada15}. The target must dissipate approximately 1.2~kW of deposited power, which can be effectively managed through beam defocusing or wobbling over the beryllium surface, supplemented if necessary by active cooling via water, air, or helium gas flow through the copper backing layer.

Once the technique is established, the same design principles can be scaled to higher-performance systems by pairing the optimized moderator/reflector geometry with research-grade cyclotrons (Table~\ref{tab:commercial_cyclotrons}). In particular, commercial models such as the IBA Cyclone IKON or the ACSI TR-30 emerge as strong candidates for achieving thermal areal densities above $10^8$\,n/cm$^2$. An even greater increase, exceeding $10^9$\,n/cm$^2$, could be realized by employing the Ta($p, xn$) reaction at proton energies around 70\,MeV, for which neutron yields are typically two orders of magnitude higher than those of the $^9$Be$(p, xn)$ reaction at proton energies below 10\,MeV~\cite{zakaleketal2025_neutron}. Such an implementation would require high-energy cyclotrons, such as the IBA Cyclone 70 listed in Table~\ref{tab:commercial_cyclotrons}, and would greatly benefit from next-generation high-current compact systems (5--10\,mA), which are still under active development~\cite{Snead23,Winklehner22}. This stepwise strategy offers a technically and economically feasible pathway for introducing the neutron-target concept to the storage-ring community while building the operational experience needed for future large-scale, high-performance dedicated installations.

\section{\label{sec:facilities} Feasibility demonstration at the CRYRING@ESR} \label{sec:CRYRING}
The primary goal of this work is to establish neutron-induced reaction measurements on short-lived nuclei. 
Therefore, the ion storage ring, into which the neutron target will be incorporated, has to fulfill several principal requirements.
Straightforwardly, it has to be coupled to a radioactive-ion beam production facility. 
The collision energy of the stored ions and thermal neutrons is defined by the kinetic energy of the ions. 
Hence, it should be able to store and manipulate ion beams at low kinetic energies ideally down to a few tens of keV/u for astrophysical applications. It is worth mentioning that most $s$-process branching nuclei measurements carried out with conventional TOF techniques are limited in the neutron-energy range up to 1-10\,keV\,\cite{Domingo25} and the inverse-kinematics methodology proposed here is thus an excellent complementary approach for these cases.

Although several projects have been discussed in the past\,\cite{Steck-2020, Litvinov-2023}, the low-energy storage ring CRYRING@ESR\,\cite{Lestinsky-2016} at the GSI Helmholtz Center for Heavy Ion Research in Darmstadt is the only suitable facility presently in operation worldwide.
Whereas proof-of-concept experiments can be demonstrated with the CRYRING@ESR, a specially designed future ion storage ring is required to boost the overall performance by several orders of magnitude (see Sec.\,\ref{sec:future}).

\subsection{The GSI facilities}
The GSI facilities are optimized for the production of radioactive ion beams employing projectile fragmentation or in-flight fission nuclear reactions.
For this purpose, the accelerator complex consisting of a universal linear accelerator (UNILAC) and a heavy-ion synchrotron (SIS18), is designed to provide any stable beam from H to $^{238}$U up to a maximum magnetic rigidity $B\rho=mv\gamma/q=18$\,Tm. $B$ and $\rho$ denote the magnetic flux density and bending radius of the magnets, and $m$, $q$, and $v$, respectively, the mass, charge, and velocity of the accelerated ions. The factor $\gamma$ stands for the relativistic Lorentz factor. 

The high-energy beams impinge on a production target installed at the entrance of the Fragment Separator (FRS)\,\cite{Geissel-1992d}, thereby producing short-lived nuclei.
The FRS can offer an inventory of (secondary) production targets, strippers and degraders enabling the preparation of a clean beam of nuclei of interest\,\cite{Geissel-1995d}.
It is important for the space-charge argumentation later that at these energies, the ions are nearly or fully ionized, which means that they have none or only a few bound electrons\,\cite{Litvinov-2011, Bosch-2013}.

The secondary ion beam is transmitted and injected into the Experimental Storage Ring (ESR)\,\cite{Franzke-1987} which is a versatile machine for manipulating the secondary ion beams\,\cite{Litvinov-2013, Steck-2020}. 
It offers fast beam cooling, further beam purification, and essentially the deceleration to lower energies\,\cite{Glorius-2023, Glorius-2023a}.
Charged-particle induced nuclear reaction experiments with an internal hydrogen or deuterium gas target have been conducted in the ESR employing decelerated stable ions \,\cite{Mei-2015, Glorius-2019, Varga-2025, Sguazzin-2025a, Sguazzin-2025b}, and very recently for the first time decelerated radioactive $^{118}$Te ions ($t_{1/2}$= 6~d)\,\cite{Dellmann-2025}.
Experience gained in these successful experiments contributed to the design of the detection systems discussed in Sec.\,\ref{Sec:Detection}.
However, the lowest center-of-mass energy reached in these experiments was about 6\,MeV, which is too high for the neutron capture reactions envisioned here.
The future perspectives to approach lower beam energies of 3-4\,MeV/u\,\cite{Sergey-priv} do not alter this conclusion.

The low-energy storage ring CRYRING has been rebuilt behind the ESR. The physics case for CRYRING@ESR focuses mainly on high-precision experiments with highly charged heavy stable ions\,\cite{Lestinsky-2016}, which defined its major characteristics.
The CRYRING is operating at extreme high vacuum (XHV) conditions of 10$^{-12}$\,mbar or better, which is decisive to achieve long storage times for low-energy beams of highly charged ions.
The CRYRING can receive ion beams from the ESR at about 10-15\,MeV/u or from a local ion source at 300\,keV/u.
The beam energy can then be adjusted inside the ring.
Electron cooling is available in the entire energy range of the CRYRING and is used to reduce and maintain the defined momentum of the beam and its spread to within 0.01\%.

Several nuclear reaction experiments of astrophysical relevance were conducted in the CRYRING employing stable beams from the local source.
For this purpose the dedicated CRYRING Array for Reaction MEasurements (CARME) has been installed at the internal gas-jet target of the CRYRING\,\cite{Marsh-2024, Marsh-2026}.
Measurements at center-of-mass energies of below 100\,keV could be achieved\,\cite{Bruno-Priv}, which is the energy range of interest here.
Experiments with beams of highly charged heavy ions provided from the ESR\,\cite{Schippers-2025, Pfaefflein-2025} demonstrated the possibility to store secondary beams, although no radioactive beam was stored yet. 

\subsection{CRYRING parameters}
The CRYRING has a circumference of 54.17\,m and a maximum magnetic rigidity of 1.44\,Tm. The injected ion beam will circulate and interact with the neutron target to be installed in a 3-m long straight section (see Figure \ref{fig:nstar}). 
The orbiting frequency is defined by the beam energy. 
For example, a $^{20}$Ne$^{q+}$ beam at 100\,keV/u circulates at 81\,kHz, while at an energy of 300\,keV/u the frequency is $\sqrt{3}$ times higher (140\,kHz).

An essential parameter is the beam lifetime, which has to be considered in addition to the radioactive decay of the ions.
The CRYRING is operating under XHV conditions to minimize beam losses due to collisions with rest gas molecules. 
Recombination reactions in the electron cooler lead to additional beam losses.
The beam lifetime and other parameters can be calculated with the dedicated online tool \texttt{Beamcalc}\,\cite{Beamcalc}. 
The lifetime of the circulating beam depends on its kinetic energy and the ionic charge state. 
The higher the charge state the larger is the recombination rate but the smaller the electron stripping rate and {\it vice versa}. 
A $^{20}$Ne$^{2+}$ beam at 300\,keV/u has a calculated lifetime of 13\,s, while the higher 6+ charge state has a lifetime of 19\,s. 
For a 1\,MeV/u $^{20}$Ne$^{6+}$ beam the lifetime can be as long as 87\,s. 

Another decisive parameter is the CRYRING space charge limit.
Calculations show that a beam containing about 10$^9$ charges can be stored\,\cite{Katayama2015}, which is in fair agreement with achieved intensities of about $2\times10^6$ $^{238}$U$^{92+}$ ions stored\,\cite{Stoehlker-priv}.
Further optimizations of the CRYRING as well as the deceleration in the ESR and beam transport are ongoing to maximize the overall stored beam intensity.
 
\subsection{Measurement cycle at the CRYRING@ESR}
For the proof-of-concept experiments, the standard GSI machine operations will be employed.
Two cycle types need to be distinguished depending whether the beam is provided by the local ion source or by the ESR. These two types differ dramatically in their duty cycles.

In the context of the envisioned running sequence, the CRYRING can be filled with the fresh beam from the local ion source basically at any time. An injected bunch at 300\,keV/u is quickly accelerated/decelerated and cooled to the energy of interest. The duration of beam preparation is at most a few seconds. At this time, the neutron target can be switched on. The duration of the measurement period depends only on the beam storage time, since the local ion source provides stable beams. 

Various neutron-induced reactions can occur. The products of ($n,\alpha)$ and ($n,p$) and other reactions except for ($n,\gamma$) have trajectories in the ring different from the stored beam. These reaction products can be immediately intercepted by particle detectors placed in the corresponding locations along the ring lattice while the unreacted beam continues to circulate. This detection principle is completely analogous to the one employed for charge-particle induced reactions studied in the ESR.

The ($n,\gamma$) reaction products require a dedicated auxiliary detection setup which is discussed in Sec.\,\ref{Sec:Detection}. For this purpose, the remaining stored beam and the ($n,\gamma$) reaction products, both with identical mean magnetic rigidities, will have to be extracted from the CRYRING. At this moment, the procedure can be repeated by refilling the ring with the fresh ions.

In the case of the beams from the ESR, the preparation of the beam at high energy and subsequent deceleration take about a minute, which is much longer than the stable ion mode.
For a radioactive beam, additional stochastic cooling and/or accumulation might be required\,\cite{Steck-2020,Glorius-2023,Leckenby-2024, Sidhu-2024}. The beam is injected into CRYRING at 10-15\,MeV/u and needs to be decelerated to the energy of interest. The remaining cycle is identical to the one of the local ion source except that the radioactive decay of the ions needs to be considered in the overall beam lifetime.

\subsection{Detection principle}
\label{Sec:Detection}
The particle orbits in a storage ring depend on their magnetic rigidities, which directly contain the orbit bending radius. Depending on the neutron-induced reaction two types of detection methods will be utilized.

\subsubsection{Radiative neutron capture: $^A$Z(n,$\gamma$)$^{A+1}$Z}
The charge $q$ is not altered in this reaction. Furthermore, due to momentum conservation, the magnetic rigidities of the parent and daughter ions are the same. Hence, the orbits of the circulating beam ($^A$Z) and the reaction products ($^{A+1}$Z) cannot be distinguished and the in-ring detectors cannot be used. The recoil of the daughter ion due to the emission of $\gamma$ radiation leads only to a small momentum broadening. 
However, both ion types have very different velocities - the (heavier) neutron capture reaction product is now circulating with a velocity that is slower by the fraction $\frac{A}{A+1}$. 

\begin{figure*}[!htb]
    \includegraphics[width=\textwidth]{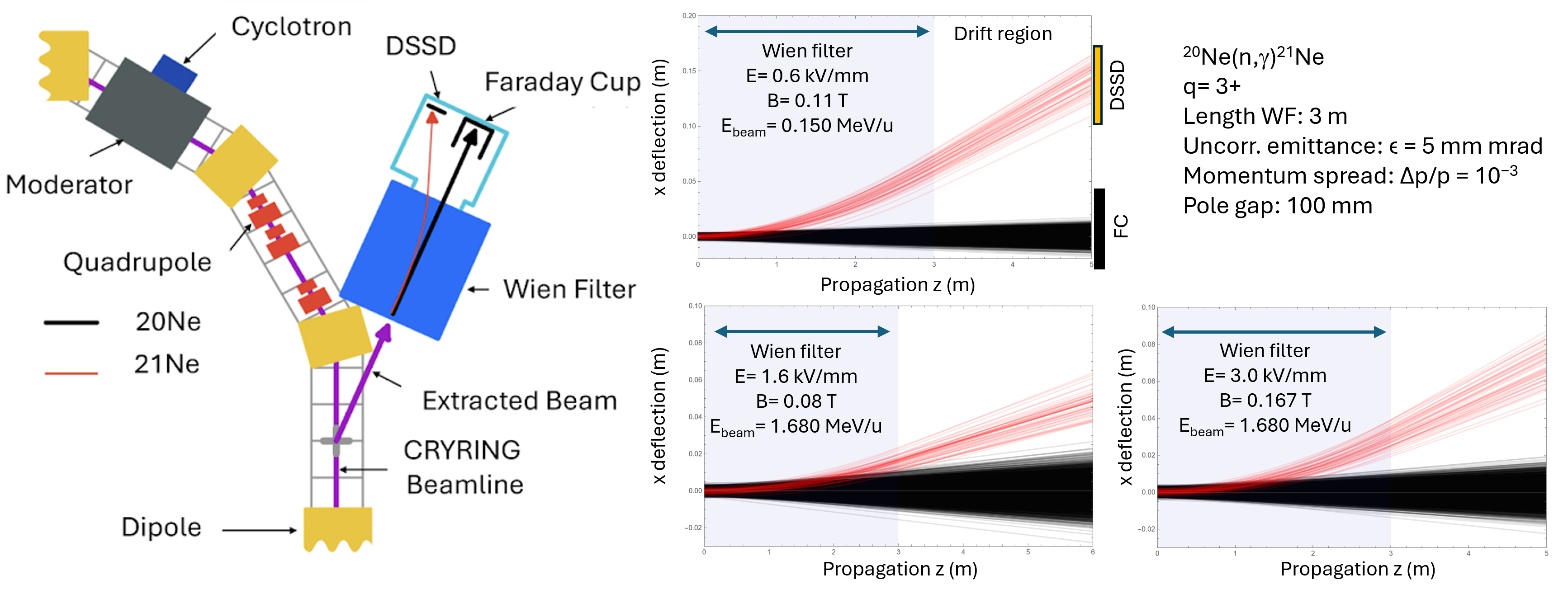}
\caption{Simulations for the separation of $^{20}$Ne$^{3+}$ (black) and $^{21}$Ne$^{3+}$ (red) with a Wien filter. (Left): Schematic drawing of
the extraction beamline with the detection setup. (Middle and right): Track simulations for 0.15 and 1.68 MeV/u with the Wien filter (blue shaded area) at various electric field strengths. }
    \label{fig:WienFilter}
\end{figure*}

To make use of this difference in velocity in the proof-of-concept experiments at the CRYRING, the beam and the reaction products will be extracted together into a Wien velocity filter (see Figure\,\ref{fig:WienFilter}). 
Exploratory simulations for the $^{20}$Ne$(n,\gamma)$$^{21}$Ne reaction at 1.68 MeV/u show that a Wien filter with a pole gap of 100 mm and an electric field strength of up to 3 kV/mm can fully separate a cooled $^{20}$Ne$^{3+}$/$^{21}$Ne$^{3+}$ beam within a drift length of at least 3\,m. This design is similar to existing Wien filters, e.g. at KoBRA at the RAON facility in South Korea\,\cite{Hwang23}. With an initial momentum spread of $\delta$p/p= 10$^{-3}$ and after a drift length (assuming an isotropic emission of a single $\gamma$-ray), the reaction products can be efficiently counted with a DSSD while the current of the unreacted beam is measured with a Faraday cup.

Various stable beams available at GSI can be employed for the validation of this new methodology for determining ($n,\gamma$) cross sections at the CRYRING. One choice could be the stable beam of $^{149}$Sm, provided either via the ESR injection or directly from a local source, if available. $^{149}$Sm has a very high and well-known radiative capture cross section, 1820(17)\,mb at $kT=30$\,keV\,\cite{kadonisv1}, and 525\,mb at $E_n=300$\,keV\,\cite{Macklin86}. Additionally, the two main existing time-of-flight measurements agree within $\sim$10\% and covered the neutron-energy range up to $E= 225$\,keV\,\cite{Wisshak93} and $E= 700$\,keV\,\cite{Macklin86}. A new TOF experiment over a broader energy range and with an improved accuracy is planned at CERN n\_TOF\,\cite{Alcayne25}.

The same measurement principle can be used for $(n,2n)$ reactions at higher energies ($E_n >10$\,MeV). However, these reactions are less interesting for nuclear astrophysics and are not discussed here. 

A more efficient design at a future dedicated facility (see e.g. Figure\,\ref{fig:TRISR}) will incorporate the Wien filter directly into the storage ring lattice. In this case only the reaction products are extracted, thus allowing a quasi-continuous measurement and enabling the beam accumulation option. Hence, a boost in luminosity by several orders of magnitude can be expected, dependent on the beam accumulation efficiency. This is discussed in Sec.~\ref{sec:TRISR} and Ref.~\cite{Pak24}.

\subsubsection{Other neutron-induced reactions: \mbox{$^A$Z(n,p)$^{A}$Z-1} and $^A$Z(n,$\alpha$)$^{A-3}$Z-2} 
These ``charge-changing" reactions are less demanding in terms of detection in the storage ring and will employ well-established methods at the ESR. This includes particle counters (e.g. DSSDs) that have to be installed directly into the CRYRING vacuum and are positioned horizontally in the beamline with movable actuators \,\cite{Glorius-2023,Najafi16, Glorius-2023a}.

In addition, non-destructive and frequency-resolved monitoring of the beam is performed with Schottky pickups \cite{Nolden11,Sanjari20}. The main task of this detector will be the determination of the number of parent ions stored, needed for the redundant luminosity determination, and their momentum distribution.

\section{\label{sec:future} Future facilities and prospects}
The CRYRING@ESR offers the unique possibility to validate the concept at an existing facility.
However, it was built for a very different research and thus has limitations that cannot be overcome. With the construction of a dedicated storage ring at an Isotope Separation On-Line (ISOL) facility\,\cite{Ravn1979} for radioactive ion beam production, several orders of magnitude gain in luminosity can be achieved.

In the ISOL method the radioactive nuclei are produced ``at rest" inside a heavier target material that is bombarded with light projectiles (protons, deuterons, or secondary neutrons) or high-energy photons (via electron-to-photon converters).

The (singly charged) ions are extracted from the target, if needed cleaned from isobaric contaminations by resonance laser ionization, mass-separated, and transported at keV-energies to the low-energy experiments (e.g. for decay spectroscopy, laser spectroscopy, mass measurements with ion traps)\,\cite{ISOLDE,Dilling14}. The extracted ion beams can be further accelerated at ISOL facilities with the help of Electron Beam Ion Sources (EBIS) that are well suited for the production of higher charge states (lower A/q ratios) for further ion selection. These post-accelerated beams could be directly injected into a connected low-energy storage ring at the energy of (astrophysical) interest for reaction experiments without any further energy adjustment. In this way, the lengthy inefficient deceleration scheme at fragmentation facilities is completely avoided (see Sec.\,\ref{sec:CRYRING}).

It is a huge advantage that the beam can be continuously accumulated in the ring, allowing to reach beam intensities close to the space charge limit in the mA range. The low atomic charge states from an ISOL beam are an asset here. Such phase-space accumulation techniques are well-established \cite{Grieser12} but the accumulation efficiency depends also on the beam lifetime. Overall, a gain of 2--3 orders of magnitude compared to the CRYRING@ESR is realistic.

However, the aforementioned accumulation scheme can only work if the velocity measurement is conducted directly inside the ring, e.g. by a Wien filter incorporated directly into the ring lattice. In this way the $(n,\gamma)$ reaction products can be removed from the ring without affecting the stored accumulated beam \cite{Pak24}.

In the following we briefly describe two ISOL-based options that may become available in the coming decades. Relevant cases of astrophysical interest are discussed later in Sec.\,\ref{sec:astro}.

\subsection{Ion Storage Ring (ISR) at CERN-ISOLDE}
At CERN ISOLDE \cite{ISOLDE}, radioactive ions are produced via spallation, fragmentation, and fission reactions induced by a proton beam from the PS-Booster impinging onto thick targets. The PS-Booster typically operates at 1.4 GeV (95\% of runs), with 2.0 GeV expected in the future. 
Singly-charged radioactive ions from the on-line mass separator are post-accelerated with HIE-ISOLDE from 30 keV to energies close to 10 MeV/u. Ions are accumulated in a Penning trap (REXTRAP) and then transported to an EBIS (REXEBIS) where their charge state can be increased.
 
\begin{figure}
    \resizebox{0.45\textwidth}{!}{\includegraphics{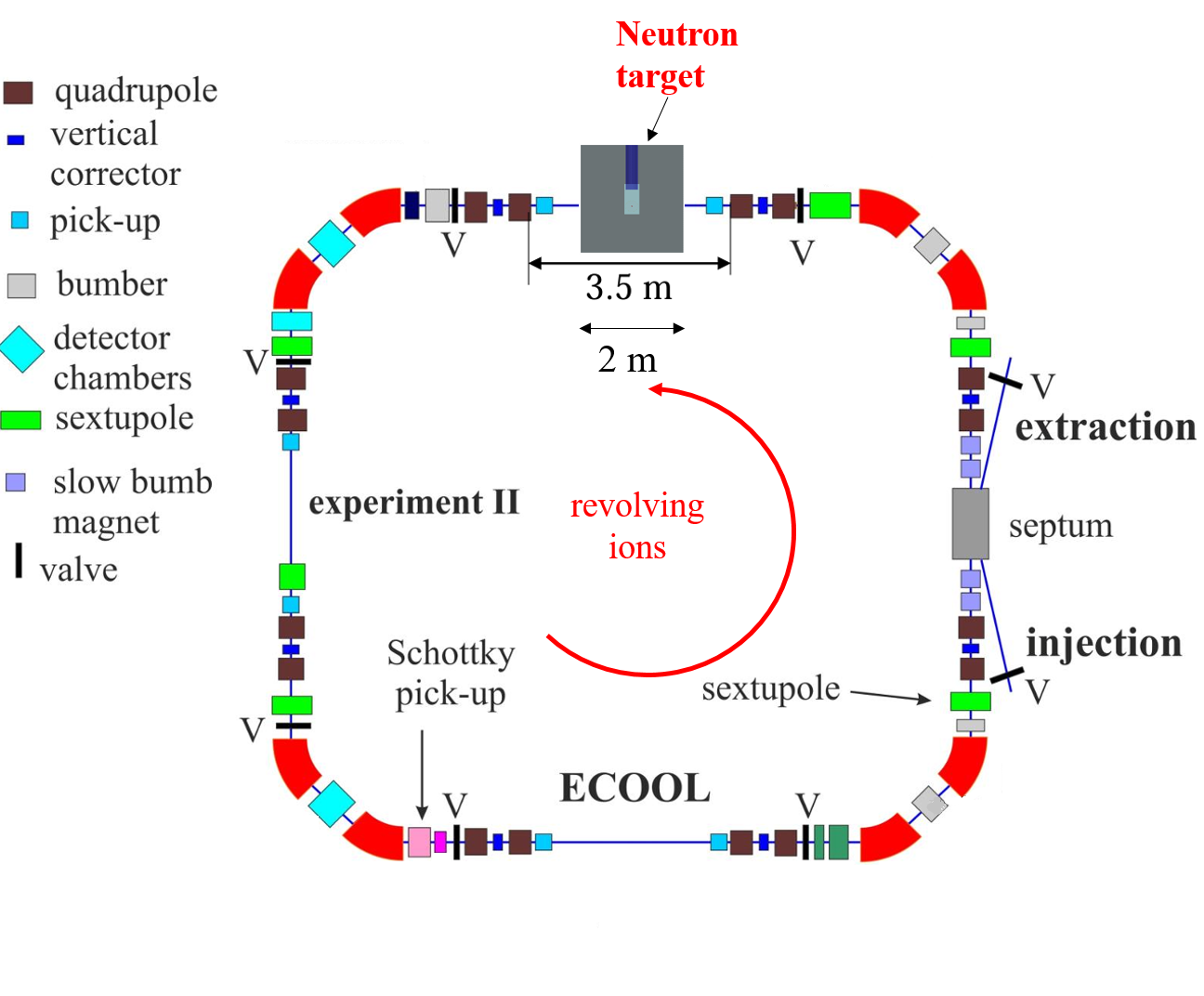}
}
\caption{Schematic drawing of the low-energy ISOLDE Ion Storage Ring (ISR) proposed in the EPIC project\,\cite{Catherall19}, to show the possible integration of the neutron target for capture experiments in inverse kinematics. Adapted from Ref.\,\cite{Catherall19}.}
\label{fig:isr}       
\end{figure}

A proposal to install the low-energy Test Storage Ring (TSR) \cite{Grieser12} at HIE-ISOLDE has been thoroughly prepared but not yet realized at CERN. A follow-up proposal considers a new dedicated Ion Storage Ring (ISR) with a circumference of 40 m \cite{Catherall19}, as illustrated in Figure\,\ref{fig:isr}. This design is optimized for multi-turn injection from HIE-ISOLDE and features electron cooling with times spanning from a few seconds for light nuclides like $^7$Be, $^{18}$F, to a few hundred milliseconds for heavy isotopes like $^{132}$Sn$^{45+}$.

The ISR design is not yet finalized and could be adapted to incorporate the neutron target and a velocity filter presented in this work.

\subsection{TRIUMF Storage Ring (TRISR) at TRIUMF-ISAC}\label{sec:TRISR}

The Isotope Separator and ACcelerator (ISAC) facility at TRIUMF, Canada, is presently the highest power radioactive ion beam facility of the ISOL type. Rare isotopes are produced by spallation, fission, and fragmentation reactions in the production targets induced by 500\,MeV proton beams of up to 100\,$\mu$A delivered by TRIUMF’s main cyclotron \cite{Dilling14}. 

A second production facility for neutron-rich nuclei via photo-induced fission, covered under the Advanced Rare IsotopE Laboratory (ARIEL) project, is presently under construction and expected to provide cleaner neutron-rich beams by 2028/29. ARIEL uses an electron linear accelerator, impinging electrons of 30\,MeV on a high-$Z$ converter material. The produced high-energy photons will induce fission reactions in the uranium targets, suppressing any other reaction channel. With the ARIEL facility, up to three radioactive beams (two from proton spallation, one from photo-fission) will be available for various experiments in the ISAC halls. 

The radioactive ions of interest are extracted from the target, mass separated, and delivered as singly-charged, low-energy beams (E$\leq$30\,keV) into the ISAC-I experimental hall. For further acceleration, higher charge states are produced by the existing Electron Cyclotron Resonance Ion Source (ECRIS) or in future with the new CANREB-EBIS. 
This allows ion beams to be further accelerated  through the radio frequency quadrupole (RFQ) and drift-tube linear (DTL) accelerators to energies of 0.15--1.8\,MeV/u. 
For nuclear structure studies the ion beams can be further accelerated with superconducting RF cavities up to 16.5\,MeV/u and delivered to experiments in the adjacent ISAC-II experimental hall.

\begin{figure}
  \resizebox{0.45\textwidth}{!}{\includegraphics{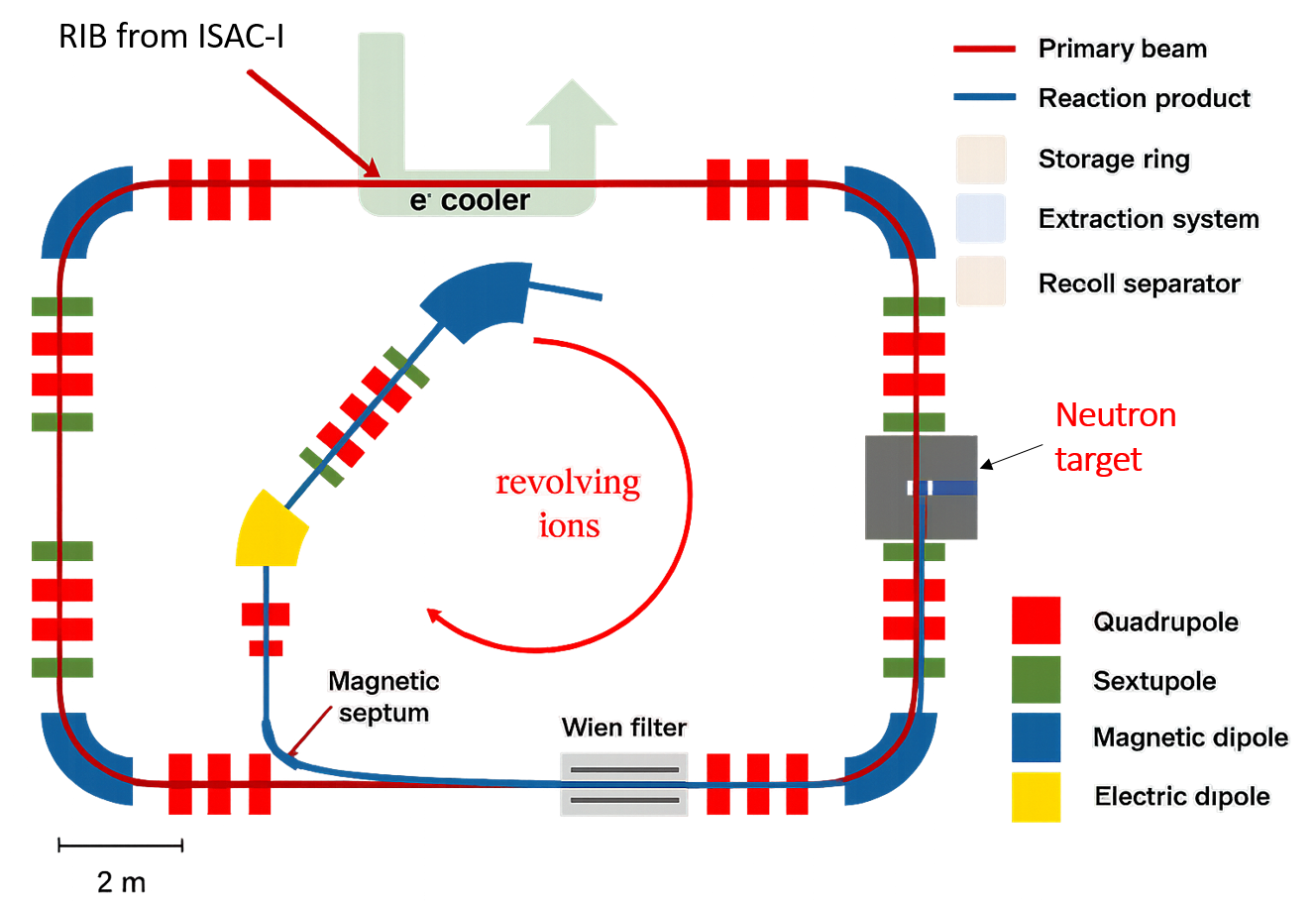} }
\caption{Drawing of TRISR, the recoil separator and the compact neutron target. Adapted from Ref. \cite{Pak24}.}
\label{fig:TRISR}       
\end{figure}

The TRIUMF Storage Ring (TRISR) project\,\cite{Dillmann23} is based on a new low-energy storage ring that would be installed in the ISAC-I experimental hall, allowing to exploit both stable beams from the offline laser ion source (OLIS) as well as intense radioactive beams from the ISAC and ARIEL target stations. The TRISR will be a storage ring of about 40–50\,m circumference, similar to the ISR foreseen at CERN-ISOLDE, covering the astrophysically interesting energy range that can be injected from the ISAC-I acceleration chain ($\simeq$ 0.15\,MeV/u up to 1.8\,MeV/u) for A/q $\leq$ 7. 

Figure\,\ref{fig:TRISR} shows a schematic view of the TRISR with the injection from the existing ISAC facility, an electron cooler, the neutron target, and the extraction beamline with a recoil separator in the center of the ring. 

The dedicated design allows the integration of a Wien filter into the ring lattice to separate reacted and unreacted beams. The reaction products are then selectively extracted into the recoil separator while the unreacted beam continues to circulate. This new concept was studied in Ref.~\cite{Pak24} based on ion-optical and particle-tracking calculations and shows that a Wien filter can be efficiently integrated without disturbing the primary circulating beam. The neutron-capture reaction products can be extracted from the ring within $\mu$s after the reaction via a magnetic (Lambertson) septum and identified with particle detectors in the focal plane of the recoil separator. With this concept,  the unreacted beam continues to circulate which is the prerequisite for the beam accumulation.
Depending on the decay half-life, production yield, and  losses of the ions of interest in the ring, several orders of magnitude higher luminosities can be achieved compared to the CRYRING concept (see Sec.\,\ref{sec:CRYRING}).

\section{\label{sec:astro} Astrophysics cases}
The neutron target design presented in Sec.\,\ref{sec:implementation} is achievable with commercially available cyclotron technology (Table~\ref{tab:commercial_cyclotrons}). It can reach areal densities of $\sim$$3\times10^6$\,n/cm$^2$ at the CRYRING proof-of-concept demonstrator (Sec.\,\ref{sec:CRYRING}) up to $\gtrsim$$10^9$\,n/cm$^2$ with higher-energy commercial models. Combined with a dedicated ISOL-based storage ring discussed in Sec.\,\ref{sec:future}, it would enable direct measurements of a large number of neutron-capture cross sections that have so far been inaccessible to conventional techniques.

The ISOL method allows access to a large number of clean, high-intensity radioactive beams. Reaction measurements will require (post-accelerated) intensities of $>$$10^{7}$\,pps for injection into the storage ring and thus will limit the range of accessible nuclei closer to stability. (Some) examples for potential neutron-capture measurements of astrophysical interest are listed in Table~\ref{tab:ISOL} together with their measured yields from extracted beams. Since TRIUMF-ISAC is about 25~years younger than CERN-ISOLDE, the yield measurements of many of these neutron-rich nuclei have not yet been performed. Some cases are listed where a value can be extrapolated by measured yields of neighboring isotopes. Close to stability, where the release times are much shorter than the decay half-lives, the yield decreases by roughly one order of magnitude per mass unit towards more neutron-rich or neutron-deficient nuclei.

It should be also noted that the ISOLDE yields are given as in-target yields, without considering release and extraction efficiencies. So the comparison with the measured ISAC yields should be taken with care. To post-accelerate the beam to the energy of astrophysical interest before injection into the storage ring, further losses of roughly in the 1--2 orders of magnitude in intensity have to be considered. This can be seen by comparison of the ISAC yields for singly-charged and highly-charged $^{134}$Cs and $^{132}$Sn. 

\begin{table*}[!htb]
\centering
\caption{Example of isotopes of astrophysical interest with their production yields at ISAC \cite{ISAC-yields} and ISOLDE \cite{Ballof2020}. The label in the first column ($s/i/r$) denotes the astrophysical process of interest.
ISOLDE yields were calculated from yield per $\mu$C as in Ref.~\cite{Ballof2020} to pps by multiplication by 1.6 assuming an average proton current of 2 $\mu$A. Note that ISOLDE values correspond to in-target yields and thus do not include the release and extraction efficiencies.  
ISAC yields are given in particles per second (pps) measured at the yield station. If yields for the respective isotope were not measured yet, short-range extrapolations are provided with a factor of 10 per mass unit, see text. Note that post-accelerated beam intensities will be lower by 1-2 orders of magnitude. Two measured yields for higher charge states of $^{132}$Sn and $^{134}$Cs are given for comparison. All other charge states are 1+.} 
\label{tab:ISOL}
\begin{tabular}{cccll} 
\hline
Process & Isotope & Half-life & Facility & Yield \\
\hline
$s$ & $^{81}$Kr & 2.3$\times$10$^5$~yr & ISOLDE & 7.0$\times$10$^6$~pps \\
$s$ & $^{85}$Kr & 10.766~yr & ISOLDE & 4.8$\times$10$^9$~pps \\
$s$ & $^{134}$Cs & 2.06~yr &  ISAC (1+) & extrapolated from $^{132}$Cs: $\approx$1.5*10$^9$~pps\\
  &   &   & ISAC (24+) & extrapolated from $^{132}$Cs: $\approx$5.5*10$^7$~pps\\
 & & & ISOLDE & $\sim$1.2$\times$10$^{10}$~pps$^{\dagger}$\\
$s$ & $^{147}$Nd & 10.981~d &  ISAC & not yet measured, $^{146,148}$Nd are stable\\
&  & &   ISOLDE & $\sim$8.6$\times$10$^6$~pps $^{\dagger}$\\
$s$ & $^{148}$Pm & 5.37~d &  ISOLDE & $\sim$8.5$\times$10$^5$~pps$^{\dagger}$\\
$s$ & $^{153}$Gd & 240.4~d & ISAC & measured: 2.2$\times$10$^{8}$~pps  \\
$s$ &  $^{160}$Tb & 72.3~d &  ISOLDE & $\sim$ 1.3$\times$10$^7$~pps$^{\dagger}$\\
$s$ & $^{170}$Tm & 128.6~d &  ISAC & extrapolated from $^{172}$Tm: $\approx$10$^9$~pps\\
 &  &   &  ISOLDE & yield $\sim$1.1$\times$10$^{10}$~pps$^{\dagger}$\\
\hline
$i$ & $^{66}$Ni & 54.6~h &  ISOLDE & 1.6$\times$10$^{8}$~pps \\
$i$ & $^{72}$Zn & 46.5~h &  ISOLDE & 1.6$\times$10$^{8}$~pps \\
$i$ & $^{75}$Ga & 126~s & ISAC & measured 5.0$\times$10$^{6}$~pps  \\
 &   &   & ISOLDE & 4.8$\times$10$^{7}$~pps \\
$i$ & $^{78}$Ge & 88~m & ISOLDE & 1.6$\times$10$^{8}$~pps \\
$i$ & $^{79}$Ge & 39~s & ISOLDE & 1.9$\times$10$^{8}$~pps \\
$i$ & $^{80}$Ge & 29.5~s & ISOLDE & 4.0$\times$10$^{6}$~pps \\
$i$ & $^{85}$Br & 2.87~m & ISOLDE & 1.2$\times$10$^{8}$~pps \\
$i$ & $^{137}$Cs & 30.08~y & ISOLDE & 2.7$\times$10$^{10}$~pps \\
$i$ & $^{141}$Ba & 18.27~m &  ISAC & extrapolated from $^{142}$Ba: $\approx$2.9$\times$10$^{10}$~pps\\
$i$ & $^{153}$Sm & 46.27~h &  ISOLDE &  1.4$\times$10$^{8}$~pps\\
\hline
$r$ & $^{132}$Sn & 39.7~s & ISAC (1+) & measured: 1.0$\times$10$^{8}$~pps\\
  &   &  & ISAC (20+) & measured: up to 5.4$\times$10$^{6}$~pps\\
$r$ & $^{132}$Sn & 39.7~s & ISOLDE & 4.8$\times$10$^{8}$~pps\\
$r$ & $^{133}$Sb & 2.34~s & ISOLDE & yield 2.1$\times$10$^{7}$~pps \\
\hline
\end{tabular}
\begin{tabular}{l}
$^{\dagger}$ Estimate from SC yields at 0.6 GeV instead of PSB at 1-1.3 GeV.
\end{tabular}
\end{table*}

In the following we discuss in more detail some astrophysically relevant cases. These cases are organized by increasing experimental challenge: from Big Bang nucleosynthesis (accessible with the CRYRING demonstrator), through $s$-process  branching points (requiring ISOL facility integration), to $i$- and $r$-process nuclei (requiring full-performance implementations with beam accumulation).

\subsection{Big Bang nucleosynthesis}
Neutron-induced reactions on light radioactive nuclei are relevant for Big Bang Nucleosynthesis (BBN) studies\,\cite{Cyburt16,Boyd10,Wiescher90}. As an example, the two reactions $^7$Be($n,p$) and $^7$Be($n,\alpha$) determine the destruction of $^7$Be in BBN and could shed light on the factor of three discrepancy between BBN models and the observed $^7$Li abundance, a long standing issue referred to as the ``cosmological lithium problem" \cite{Cyb16,Fields23}. 

Direct measurements of the $^7$Be($n,p$)$^7$Li \cite{Damone18} and $^7$Be($n,\alpha$)$^4$He reactions were recently performed at CERN n\_TOF, from thermal neutron energies up to 325 keV and 25 keV, respectively \cite{Barbagallo16}. The main limitation of these (and previous) measurements was the large $^7$Be sample activity ($t_{1/2}$= 53.1\,d) of up to 36\,GBq, which hindered the measurement in the full energy range of astrophysical relevance. 

In this respect, future experiments at storage rings could help to fully cover the energy window relevant for BBN and circumvent any issues with the sample radioactivity, thus extending previous measurements even up to the few MeV range.  Such studies could be best tackled at ISOLDE, where the yield for $^7$Be is $2.8\times10^{10}$\,pps (for comparison, the highest ISAC yield is ``only" $4.4\times10^{9}$\,pps).
The expected half-life of the electron-capture of $^7$Be$^{3+}$ (with only one electron left) is about 100\,d -- this measurement alone is interesting for the solar neutrino flux estimation\,\cite{Langanke, Grieser12}. 

Only a few $10^6$ -- $10^7$ stored $^7$Be$^{3+}$ ions  in a low-energy storage ring are required for an experiment in inverse kinematics with the proposed neutron target. Taking into account the transmission efficiency and space charge limitations of REXTRAP, the estimated intensity at the entrance of the storage ring is about $10^8 - 10^9$ $^7$Be$^{3+}$ ions per second. This beam intensity is so high that an accumulation scheme may not be required. 

In combination with our expected areal neutron density of 10$^9$ n/cm$^2$ and a revolution frequency of 140\,kHz at 300\,keV/u, where the cross section is $\sim$4~mb \cite{Damone18}, this gives a luminosity of 10$^{22}$--10$^{23}$ cm$^{-2}$s$^{-1}$ or, equivalently, about 200--2000 counts/h. Such a detection rate would be sufficient to scan the full energy range up to several MeV in 100\,keV steps within just a few days of experiment.


\subsection{Experiments along the valley of $\beta$-stability: The $s$-process branching points}

One of the long-standing challenges in the study of the $s$-process concerns the measurement of ($n,\gamma$) cross sections of several radioactive branching points \cite{Kaeppeler11}. Branching point nuclei are isotopes whose decay half-life are in the order of the neutron capture time scale, allowing the $s$-process reaction path to break into two ``branches" depending on the neutron density and temperature. Due to this, their neutron capture cross sections and low-lying nuclear structure are important factors for the indirect measurement of stellar temperatures or neutron densities. Despite this importance, many of the shorter-lived branching nuclei (see Table~\ref{tab:branchings}) could  thus far not be measured with direct methods due to limitations in experimental techniques. 

Many other relevant branching nuclei could be only accessed in the low neutron-energy range of up to a few keV\,\cite{Dillmann23,Domingo23b,Domingo25}. A review on the astrophysical relevance of the $s$-process branching points and the related cross section uncertainties can be found in Ref.~\cite{Bisterzo15}. 

\begin{table}[htbp]
\centering
\caption{Main unmeasured $s$-process branching isotopes. $T$ stands for temperature and $\rho_n$ for neutron density. The index $(r)$ indicates refractory element not accessible at ISOL facilities.}
\label{tab:branchings}
\begin{tabular}{cccc} 
\hline
\textbf{Isotope} & \textbf{Half-life}  & \textbf{Comments}\\
\hline
$^{81}$Kr & 2.29$\times$10$^5$~yr & $T$ in massive stars\\
$^{85}$Kr & 10.73~yr & $\rho_n$ in massive stars\\
$^{95}$Zr$^{(r)}$ & 64.02~d & $\rho_n$ in AGBs\\
$^{134}$Cs & 2.0652~yr & $T$ in AGBs\\
$^{147}$Nd & 10.981~d & $\rho_n$ in AGBs\\
$^{148}$Pm & 5~d& impacts the $s$-only $^{148}$Sm/$^{150}$Sm\\
$^{154}$Eu & 8.593~yr & $T$ and $\rho_n$ in AGBs\\
$^{153}$Gd & 0.658~yr &  $T$ and $\rho_n$ in AGBs\\
$^{160}$Tb & 0.198~yr & $\rho_n$ in AGBs\\
$^{170}$Tm & 0.352~yr & $\rho_n$ in AGBs\\
$^{179}$Ta$^{(r)}$ & 1.82~yr & origin of rarest stable $^{180}$Ta\\
$^{185}$W$^{(r)}$ &  0.206~yr & $T$ and $\rho_n$ in AGBs\\
$^{186}$Re$^{(r)}$ & 3.72~d & $^{187}$Os/$^{187}$Re cosmic clock\\
$^{192}$Ir$^{(r)}$ & 73.826~d & $s$-only $^{192}$Pt anomaly (-20\%)\\
\hline
\end{tabular}
\end{table}

Nine of these unstable species have half-lives shorter than 1~yr, which makes them inaccessible for direct measurements via state-of-the-art activation or time-of-flight techniques. For the other five longer-lived isotopes it is very difficult and expensive to produce samples with sufficient enrichment and quantity since around 10$^{15}$ atoms are required for a direct measurement. With a yield of 10$^8$ pps at an ISOL facility, the production of such a sample would require 116~days of uninterrupted collection time. 

One major achievement of CERN n\_TOF was the measurement of the neutron capture cross sections of seven $s$-process branching points over the last 25 years \cite{Domingo25}: $^{63}$Ni (100.8\,yr), $^{93}$Zr (1.61\,Myr), $^{94}$Nb (20.4\,kyr), $^{79}$Se (0.327\,Myr), $^{151}$Sm (94.6\,yr), $^{171}$Tm (1.92\,yr) and $^{204}$Tl (3.78\,yr). However, only $^{151}$Sm could be measured in the full energy range of interest for nucleosynthesis (1\,eV - 100\,keV). The measurement of the other six species was limited to neutron energies of only a few keV\,\cite{Domingo25}. Additional measurements at higher energies up to a few MeV with the storage ring method would be very important for a more comprehensive astrophysical interpretation.

We consider here, as an example, the possibility of using the CRYRING for some of these measurements. Given the high intensities for primary Kr beams of about $2\times10^{10}$ particles per spill, the measurement of both branchings $^{81,85}$Kr($n,\gamma$) could be feasible in an experiment combining the FRS with the storage rings ESR and CRYRING. The FRS could be used as a high-resolution separator to select a secondary beam of $^{85}$Kr (or $^{81}$Kr). This beam could be accumulated to $10^7 - 10^8$ ions/s in the ESR with injections every few seconds, and then decelerated and transported to the CRYRING. Assuming a revolution frequency of 140\,kHz for a 300\,keV/u beam and an estimated cross section of about 30\,mb, with the expected areal neutron density of 10$^9$\,n/cm$^2$ one would expect 10’s of counts per day, sufficient to perform a complete measurement in a few days of experiment at the chosen energy.

ISOL beams (see Table~\ref{tab:ISOL}) could provide sufficient intensities for direct neutron-capture measurements on most of the yet unmeasured lighter $s$-process waiting points. 
However the extraction of ions out of the target material at such facilities is highly chemistry-dependent.
Many of the non-refractory $s$-process branching isotopes in Table \ref{tab:branchings} would be well accessible for ($n,\gamma$) measurements using future low-energy storage rings at ISOL facilities. 
A very straightforward case is $^{134}$Cs, which can be easily produced in high quantities (yields $>$10$^{8}$~pps) at both ISOL facilities discussed here. For  refractory elements with high melting points like Zr, Ta, W, Re, and Ir, the use of in-flight RIB facilities will remain unavoidable.

\subsection{Experiments for the $i$- and $r$-process}
Both the intermediate ($i$)- \cite{Cowan77} and the rapid ($r$) neutron capture processes \cite{BBFH} are characterized by much higher neutron densities than the $s$-process, and their reaction paths lie on the more neutron-rich side of the nuclear chart. These processes involve hundreds and thousands of unstable neutron-rich nuclei, respectively. Because of their short half-lives, they are completely out of reach for conventional methods, and only indirect techniques like surrogate reactions\,\cite{Escher12,Jurado21} or the $\beta$-Oslo method\,\cite{Spyrou2014,Liddick2016,Larsen2019,Liddick2019} can be employed in some cases to constrain the cross sections. 

As for all three neutron-driven nucleosynthesis mechanisms ($s$, $i$, and $r$ process), neutron-capture rates play a fundamental role in network calculations to model abundances of the corresponding stellar environments\,\cite{Arcones23}. Some of the most debated $i$-process challenges are puzzling elemental abundance ratios observed in metal-poor stars like HD94028\,\cite{Roederer2016}. As reported in the sensitivity study of Ref.\,\cite{McKay2020}, a proper interpretation of the observed abundances in this star would require ($n,\gamma$) cross sections for the short-lived nuclei $^{66}$Ni, $^{69}$Cu, $^{72}$Zn, $^{75}$Ga, $^{78}$Ge, $^{79}$Ge, $^{80}$Ge, $^{84}$Se, and $^{85}$Br.

Other sensitivity studies focus on the possibility of the $i$-process as a plausible explanation for the abundance pattern observed in CEMP (Carbon-Enhanced Metal-Poor) stars\,\cite{Hampel2016}. For this, recent sensitivity studies\,\cite{Choplin21} identified that a better knowledge of neutron-capture cross sections for $^{137}$Cs, $^{144}$Ce, $^{153}$Sm,$^{155}$Sm, and $^{151}$Nd are required.

With the exception of $^{137}$Cs and $^{144}$Ce, none of these nuclei of interest for the $i$-process can be accessed presently for neutron-capture experiments with conventional direct techniques. However, the short half-lives represent no limitation for experiments in inverse kinematics, provided that sufficiently intense secondary beams can be produced at a RIB facility and accumulated in the storage ring. 

A few examples of nuclides that can be efficiently produced at ISOL facilities are listed in Table\,\ref{tab:ISOL}.  Some of them could be also studied via in-flight fragmentation, but the inefficient duty cycle including the deceleration from higher primary/secondary beam energies to very low energies of several 100\,keV may pose constraints to perform the neutron-capture experiments at a storage ring. The renewed interest in the $i$-process will identify additional candidates for measurements in the coming years.

For $r$-process nucleosynthesis studies, comprehensive data surveys are essential, far beyond isolated measurements on single nuclei \cite{Surman08,Arcones23}. Neutron captures reactions play a decisive role in shaping the final abundance patterns, defining the positions and shapes of the $r$-process peaks at $A\approx$ 80, 130, and 195, and driving the formation of the rare-earth peak at $A\approx$160. Although most models along the $r$-process path assume an ($n,\gamma$)$\rightleftharpoons$($\gamma,n$) equilibrium (where single neutron captures do not play a big role), neutron capture rates become particularly critical during the pre-equilibrium phase and the subsequent freeze-out, where they directly influence the pattern of the abundance distribution. 
Earlier sensitivity studies showed that cross sections of nuclei around the $N$=82 and 126 shell closures at $A$$\approx$130 and 195 have the highest impact on the abundances, for any of the discussed explosive astrophysical scenarios ~\cite{Surman14,Mumpower16}. For the finetuning of reaction models, on the other hand, systematic measurements along isotopic chains are extremely valuable.

\section{\label{sec:summary} Summary and outlook}
This work introduced a novel concept for a free-neutron target optimized for the integration into a low-energy storage ring. This would enable for the first time direct measurements of neutron-induced reactions on radioactive ions in inverse kinematics. 

The proposed neutron target design integrates four subsystems to maximize the thermal neutron density within a small interaction volume inside the storage-ring beam pipe, where radioactive ions circulate at high frequencies (100\,kHz--1\,MHz). The system combines a compact cyclotron-driven $^9$Be($p,xn$) neutron source with an optimized moderator assembly consisting of a primary moderator core (D$_2$O or BeO), a graphite reflector shell, and a cryogenic liquid hydrogen moderator at 20\,K positioned directly around the ion beam path. This configuration, optimized through extensive Monte Carlo simulations, achieves thermal neutron areal densities normalized to the primary neutron source strength of $\sim 15 \times 10^{-7}$\,n/cm$^2$/prim. Overall, this approach provides a simpler and more cost-effective alternative to neutron-target concepts based on reactor or spallation sources, while offering a modular solution that can be readily integrated into future low-energy storage rings without major infrastructure modifications.

In this work we have also introduced a proof-of-concept setup at the CRYRING storage-ring facility of GSI Darmstadt utilizing a commercially available proton cyclotron (130 $\mu$A, 9.2 MeV), which could deliver a total areal neutron density of 3.4$\times$10$^6$ n/cm$^2$. This setup would serve as a minimum viable platform for initial demonstration and validation experiments on various neutron-induced reactions, including radiative capture.

Other commercially available cyclotrons, still compact and compatible with a storage-ring facility, would allow one to reach proton currents of $\sim$1.6~mA, thus leading to areal densities of 1.3$\times$10$^8$ n/cm$^2$ (see Table~\ref{tab:commercial_cyclotrons}). 
Recent developments of compact isochronous cyclotrons \cite{Snead23,Winklehner22} will allow a one order of magnitude increase in current within the coming decade, up to $\sim$10~mA, which together with a beryllium or a tantalum target translates into areal neutron densities well above 10$^9$ n/cm$^2$.

In combination with customized low-energy storage rings that can accumulate 10$^{9}$ ions, reaction luminosities of 1.4$\times$10$^{23}$\,cm$^{-2}$s$^{-1}$ can be achieved. Assuming 100\% detection efficiency, the daily count rate would be $\sim$12$\times\sigma$\,~events/day (with $\sigma$ being the cross section of interest in mb). This means that reaction cross sections as low as a few mb could be measured in a few days of experiment at $\lesssim$10\% statistical uncertainty level. 

Future developments of the neutron target concept will focus on several key areas: further optimization of the target geometry, implementation of superconducting cyclotron upgrades to boost performance, and the integration with existing and proposed storage rings (e.g., TRISR at TRIUMF, ISR at ISOLDE). These developments would benefit enormously from improved nuclear data, particularly neutron production yields, angular distributions, and energy spectra for the ${}^{9}$Be$(p,xn)$ reaction in the 5--30\,MeV and the Ta$(p,xn)$ reaction in the 50--80\,MeV range. Enhanced characterization of these nuclear reactions would enable more precise Monte Carlo simulations and improved target optimizations, ultimately leading to higher neutron densities and better performance predictions for next-generation facilities.

The proposed setup opens new opportunities for measuring a large number of ($n,\gamma$) reactions on unstable isotopes relevant to astrophysics, which are completely out of reach with state-of-the-art time-of-flight or activation techniques. Key nuclei for these future measurements include a large list of branching points in the $s$-process, important reactions in Big Bang nucleosynthesis, and more exotic isotopes of relevance for nucleosynthesis in the $i$- and $r$-processes. 

Exploring the vast \textit{Terra Incognita} of the ($n,\gamma$) domain on both sides of the $\beta$-stability valley with extensive new surveys will not only constrain and refine theoretical models, but also make their extrapolations to nuclei further from stability significantly more reliable. In this context, the synergy between ISOL and in-flight fragmentation facilities, when combined with tailored low-energy storage rings and advanced neutron target designs, could deliver unprecedented experimental access to a wide range neutron-induced reactions on unstable neutron-rich nuclei. These breakthroughs would provide the long-sought benchmarks required to transform our understanding of the $r$-process and the cosmic origin of the heaviest elements.

\section*{Acknowledgments}
The authors would like to thank Josep Sempau at Polytechnical University of Catalonia and Jos\'e Manuel Quesada at the University of Seville for fruitful suggestions on the implementation of the particle density estimator. 
Many people have contributed with discussions or previous ideas in one way or another, and the authors would like to thank Pierre Cesar, Maeve Cockshutt, Barry Davids, Luis Mario Fraile,  Jan Glorius, Manfred Grieser, Chris Griffin, Ion Ladarescu, Guy Leckenby, Michael Lestinsky, Sergey Litvinov, Nil Mont-Geli, Rene Reifarth, and Chris Ruiz. We particularly like to remember Markus Steck, an irreplaceable expert in heavy-ion storage rings, who suddenly passed away in 2025.

This research was in part supported by AEI/10.13039/501100011033 under grants PID2022-138297NB-C21 and Severo Ochoa CEX\-2023-001\-292\--S, by the project ``NRW-FAIR", a part of the program ``Netzwerke 2021", an initiative of the Ministry of Culture and Science of the State of North Rhine-Westphalia, and by the Canadian Natural Sciences and Engineering Research Council (NSERC) via the grants SAPIN 2019-00030 and 2025-00033.
TRIUMF receives federal funding via a contribution agreement with the National Research Council of Canada (NRC). We also acknowledge the use of AI-based language tools to assist with grammar and linguistic refinement of the text.
\input{appendixI}

\bibliography{bibliography}

\end{document}

%% file: appendixI.tex
\appendix
\section{Derivation of the Particle Density Estimator}
\label{app:density-estimator}

This appendix presents the derivation of the particle density estimator normalized to source activity based on the track-length estimator, readily enabling implementation in Monte Carlo transport codes, in particular Geant4-based simulations.

\subsection{Mathematical framework}

Consider a radiation field within a volume $V$ discretized into cells $V_j$ such that $V = \sum_j V_j$. Particles are categorized by velocity groups $\mathcal{G}_i = [v_i, v_{i+1})$. The following notation is employed:

\begin{center}
\footnotesize
\begin{tabular}{lll}
\hline
Symbol & Description & Units \\
\hline
$n_{j,i}$ & Particle density (cell $j$, group $i$) & cm$^{-3}$ \\
$\phi_{j,i}$ & Fluence rate (cell $j$, group $i$) & cm$^{-2}$s$^{-1}$ \\
$\Phi_{j,i}$ & Time-integrated fluence & cm$^{-2}$ \\
$I$ & Source strength & s$^{-1}$ \\
$V_j$ & Volume of cell $j$ & cm$^3$ \\
$\bar{v}_i$ & Representative velocity of group $i$ & cm/s \\
$\ell_m$ & Track length of segment $m$ & cm \\
$S$ & Total source particles simulated & -- \\
$\mathcal{T}_{j,i}$ & Track segments in cell $j$, group $i$ & -- \\
\hline
\end{tabular}
\end{center}

The fundamental relationship between fluence rate and particle density for particles with velocities in group $\mathcal{G}_i$ is:
\begin{equation}
\phi_{j,i} = n_{j,i} \cdot \bar{v}_i
\label{eq:fluence-density}
\end{equation}

The radiation field originates from a particle source with strength $I = dN/dt$, where $N$ represents the cumulative number of primary particles emitted by the source. Under steady-state conditions, the particle population in cell $j$ for velocity group $i$ is:
\begin{equation}
N_{j,i} = n_{j,i} \cdot V_j
\label{eq:population}
\end{equation}

\subsection{Track-length estimator}

In Monte Carlo transport, the time-integrated fluence for velocity group $i$ in cell $j$ is estimated using the track-length method:
\begin{equation}
\Phi_{j,i} = \frac{1}{V_j} \sum_{m \in \mathcal{T}_{j,i}} \ell_m
\label{eq:track-length}
\end{equation}

where $\mathcal{T}_{j,i}$ denotes all particle track segments in cell $j$ with velocities in group $\mathcal{G}_i$, and $\ell_m$ is the track length of segment $m$. The fluence rate can then be expressed as $\phi_{j,i} = \Phi_{j,i}/\Delta t$, where $\Delta t$ is an arbitrary time interval.

Combining Equations~\eqref{eq:fluence-density} and \eqref{eq:track-length}, the time-independent population becomes:
\begin{equation}
N_{j,i} = \frac{\Phi_{j,i}}{\Delta t} \cdot \frac{V_j}{\bar{v}_i} = \frac{1}{\Delta t \cdot \bar{v}_i} \sum_{m \in \mathcal{T}_{j,i}} \ell_m
\label{eq:population-fluence}
\end{equation}

For a Monte Carlo simulation with $S$ primary source particles, the source strength is given by $I = S/\Delta t$. Normalizing the population by the source strength yields:
\begin{equation}
\frac{N_{j,i}}{I} = \frac{N_{j,i}}{S/\Delta t} = \frac{1}{S \cdot \bar{v}_i} \sum_{m \in \mathcal{T}_{j,i}} \ell_m
\label{eq:normalized-population}
\end{equation}

\subsection{Final particle density estimator and implementation}

Dividing Equation~\eqref{eq:normalized-population} by the cell volume $V_j$ and using Equation~\eqref{eq:population}, we obtain the particle density estimator normalized to the source strength ($\tilde{n}_{j,i}$):

\begin{equation}
\tilde{n}_{j,i}=\frac{n_{j,i}}{I} = \frac{1}{S \cdot \bar{v}_i \cdot V_j} \sum_{m \in \mathcal{T}_{j,i}} \ell_m
\label{eq:final-estimator}
\end{equation}

In Monte Carlo transport codes, Equation~\eqref{eq:final-estimator} is implemented by accumulating track-length contributions during particle transport. For each particle track segment $m$ in cell $j$ with velocity in group $\mathcal{G}_i$, the contribution $\ell_m/ V_j$ is added to the estimator. Later, the estimator is normalized by the term $1/(S \cdot \bar{v}_i)$. For neutron transport applications, velocity groups typically correspond to energy groups, with ${v}_i = \sqrt{2E_i/m_n}$ where $E_i$ is the group's representative energy and $m_n$ is the neutron mass. Moreover, $\bar{v}_i$ represents the average velocity in group $i$. The normalization factor $1/S$ ensures that results are independent of the number of simulated histories, facilitating comparison between simulations.